\documentclass[preprint]{aastex}
\begin{document}
\title{The Sunyaev Zel'dovich effect: simulations and observations}
\author{Zhang, Pengjie}
\affil{Department of Astronomy \& Astrophysics, University of Toronto,
M5S 3H8, Canada} 
\email{zhangpj@cita.utoronto.ca}
\author{Pen, Ue-Li}
\affil{Canadian Institute for Theoretical Astrophysics, University of
Toronto, M5S 3H8, Canada}
\email{pen@cita.utoronto.ca}
\author{Wang, Benjamin}
\affil{Physics and Astronomy Department, University of British
Columbia, Canada}

\begin{abstract}
The Sunyaev Zel'dovich effect (SZ effect) is a complete probe
of ionized baryons, the majority of which are likely hiding in the
intergalactic medium.  We ran a $512^3$ $\Lambda$CDM simulation using a
moving mesh hydro code to compute the statistics of the thermal and
kinetic SZ effect such as the power spectra and measures of
non-Gaussianity. The thermal SZ power spectrum has a very broad 
peak at multipole $l\sim 2000-10^4$ with temperature fluctuations
$\Delta T \sim 15\mu$K. The power spectrum is consistent with
available observations and suggests a high $\sigma_8\simeq 1.0$ and a
possible role of non-gravitational heating. The non-Gaussianity is
significant and  increases the cosmic variance of the power spectrum
by a factor of $\sim 5$ for $l<6000$. 

We explore optimal driftscan survey strategies for the AMIBA CMB
interferometer and their dependence on cosmology.  For SZ power spectrum
estimation, we find that the optimal sky coverage for a $1000$ hours of
integration time is several hundred square degrees. One achieves an
accuracy better than $40\%$ in the SZ measurement of power spectrum and
an accuracy better than $20\%$ in the cross correlation with Sloan galaxies for
$2000<l<5000$.   For cluster searches, the optimal scan rate is around
$280$ hours per 
square degree with a cluster detection rate 1 every $7$ hours, allowing
for a false positive rate of $20\%$ and better than $30\%$ accuracy in the
cluster SZ distribution function  measurement.

\end{abstract}
\section{Introduction}

Big bang nucleosynthesis (BBN) and cosmic microwave background (CMB)
experiments such as  Boomerang \citep{Netterfield01} and DASI \citep{Pryke01}
predict that the ordinary baryonic matter accounts for about $5\%$ of
the total matter in the universe. But the directly observed
components, 
such as stars, interstellar medium and intracluster gas, only
account for  a few percent of this baryon budget, while more than $90\%$
baryons have escaped direct detections
\citep{Persic92,Fukugita98}. This is known as the missing baryon
problem.  The missing baryons are believed to be in the form of the
intergalactic medium (IGM) and have been difficult to detect
directly. To understand their state, such as density, temperature,
peculiar velocity and metalicity, stands as a major challenge to both
observation and theory, and is crucial to understand the thermal
history of the universe and galaxy formation.

The IGM has various direct observable tracers. (1) Neutral hydrogen
absorbs background light of quasars and produces the
Lyman-$\alpha$ forest. (2) Ionized electrons have thermal and
peculiar motions and are capable of scattering CMB photons and
generate secondary CMB temperature fluctuations, which are known as the
thermal and kinetic Sunyaev Zel'dovich effects (SZ effect),
respectively. Their precision measurements are becoming routinely
available with the devoted CMB experiments such as AMIBA
\citep{AMIBA} and SZA (Sunyaev Zel'dovich array).  (3) Ionized
electrons and protons interact with 
each other and emit X-rays through thermal bremsstrahlung and
contribute to the soft X-ray ($0.5-2$ keV) background (XRB).  Several
other tracers have been proposed, including X-ray absorption
techniques \citep{Perna98}, but they depend strongly on chemical
compositions, and 
would be difficult to associate with direct IGM properties. These
tracers probe different IGM phases and help to extract the IGM
state. For example, the XRB flux upper limits place constrain on the
gas clumpyness and suggest a potentially strong role of feedback \citep{Pen99}.

Among these tracers, the SZ effect is a particularly powerful IGM
probe. (1) It provides a complete sample of the intergalactic medium.
All free electrons participate in Thomson scattering and contribute to
the SZ effect. The Thomson optical depth from the epoch of
reionization $z\sim 10$ to the present is $\tau \sim 0.1$, which means
that about $10\%$ of CMB photons have been scattered by
electrons. Since the number of CMB photons is much larger than the
baryon number and the ionization fraction of our universe is larger
than $99\%$, most baryons contribute to the SZ effect.
Compton scattering does not depend on redshift and is not affected by
distance or the expansion of the universe. So, the SZ effect can probe
the distant universe. Furthermore, the SZ effect does not strongly
depend on gas density and probes a large dynamic range of baryon
fluctuations. In 
contrast, the X-ray emission measure depends on density squared, and
primarily probes the densest IGM regions at low redshift.  The
Lyman-$\alpha$ forest probes the neutral IGM, which in ionization
equilibrium depends also on the square of the baryon density.  The
neutral fractions only accounts for a tiny fraction of total baryons. 
Extrapolations over many orders of magnitude are required to
understand the bulk of the baryons.  (2) It is straightforward to
understand.  The simple dependence of the SZ
effect on the ionized gas pressure or momentum does not put as stringent
requirement on simulation resolution, nor as accurate an understanding
of the gas state such as metalicity, temperature, velocity and ionization
equilibrium, as X-rays and Lyman-$\alpha$ properties do.  Pressure is the total
thermal energy content of the gas, and finite volume flux conservative
simulations are particular amenable to its modelling.  It is also
easier to model analytically. (3) It has strong 
observational potential.  In this paper we will primarily consider the
thermal SZ effect, which is easier to observe.  Its unique dependence
on  frequency allows us in principle to disentangle the SZ effect from the
contamination of the primary CMB and other noise
sources \citep{Cooray00b}. Hereafter, if not otherwise specified, the SZ effect
always means the thermal SZ effect.

In SZ observations, all redshifts are entangled and smear some key
information of the intervening IGM: its spatial distribution and time
evolution. These properties around redshift $z\sim 1$ are sensitive to
many cosmological parameters and thus potentially promising to break
degeneracies in cosmologies from the primary CMB experiments. One can
hope to resolve the equation of the state of the dark energy. We are
currently performing simulations for different cosmologies, which are
degenerate in the primary CMB, to test the potential of such a procedure
\citep{Zhang02}. If combined with other observations, the SZ effect
will become even more powerful. For example, with the aid of a galaxy
photometric redshift survey, the evolution of the 3D gas pressure
power spectrum and its cross correlation with the galaxy distribution
can be extracted \citep{Zhang01a}. 

The above analysis depends on a detailed quantitative theoretical
understanding of the SZ statistics.  Analytical approaches considered
in the past strongly depend on various assumptions.  The
Press-Schechter approach as adopted by
\citet{Cole88,Makino93,Atrio99,Komatsu99,Cooray00,Molnar00} requires ad
hoc models for the gas profile in halos, whose shape and evolution
are uncertain. The hierarchical method as proposed by
\citet{Zhang01a} strongly depends on the gas-dark matter correlation,
which is also poorly understood. These 
estimates can be significantly improved by high-resolution
simulations.  But past simulation
results \citep{Scaramella93,daSilva00,Refregier00, Seljak01,Springel01}
disagreed on both amplitude and shape of the SZ power spectrum
(see \citet{Springel01} for a recent discussion). Simulation
resolution may play a key role in this discrepancy
since the SZ effect is sensitive to small structures, as discussed by
\citet{Seljak01}.  Differences between code algorithms may also cause
part of the discrepancy. To address these problems, we ran the largest
adaptive SZ simulation to date, a direct $512^3$ moving mesh hydro (MMH) code
\citep{Pen98a} simulation,  to reinvestigate the SZ statistics.
For a better understanding of numerical and code issues 
in the SZ effect, our group is carrying out a series of
simulations with identical cosmological parameters and initial
condition but using different codes with different resolutions
\citep{Codes}.

Apart from simulation issues, a problem that has received little
attention is analysis strategies for SZ data.  Several
interferometric arrays such as AMIBA and SZA are under construction, but
detailed data analysis models are still in their infancy.  With our
simulated SZ maps, we can estimate the sensitivity and accuracy of SZ
observations.  Given a sky scan rate, how accurately can the SZ power
spectrum be measured? How many cluster can be found?  How
accurately can the SZ decrement be determined for individual clusters?
What are optimal strategies for these measurements?  In this paper, we
will take AMIBA as our target to address these questions.

This paper is organized as follows: in \S \ref{sec:application}, we
describe the SZ effect, its 
statistics and our method to analyze these statistics.  We then
present our simulation 
results of these statistics and possible constrain from observations
(\S\ref{sec:simulation}). In 
\S\ref{sec:AMIBA} we simulate AMIBA driftscan observations to
estimate the accuracy of AMIBA measurement of these statistics.  We
conclude in \S \ref{sec:conclusion}.

\section{Formulation and Background Review}
\label{sec:application}
\begin{figure}
\plotone{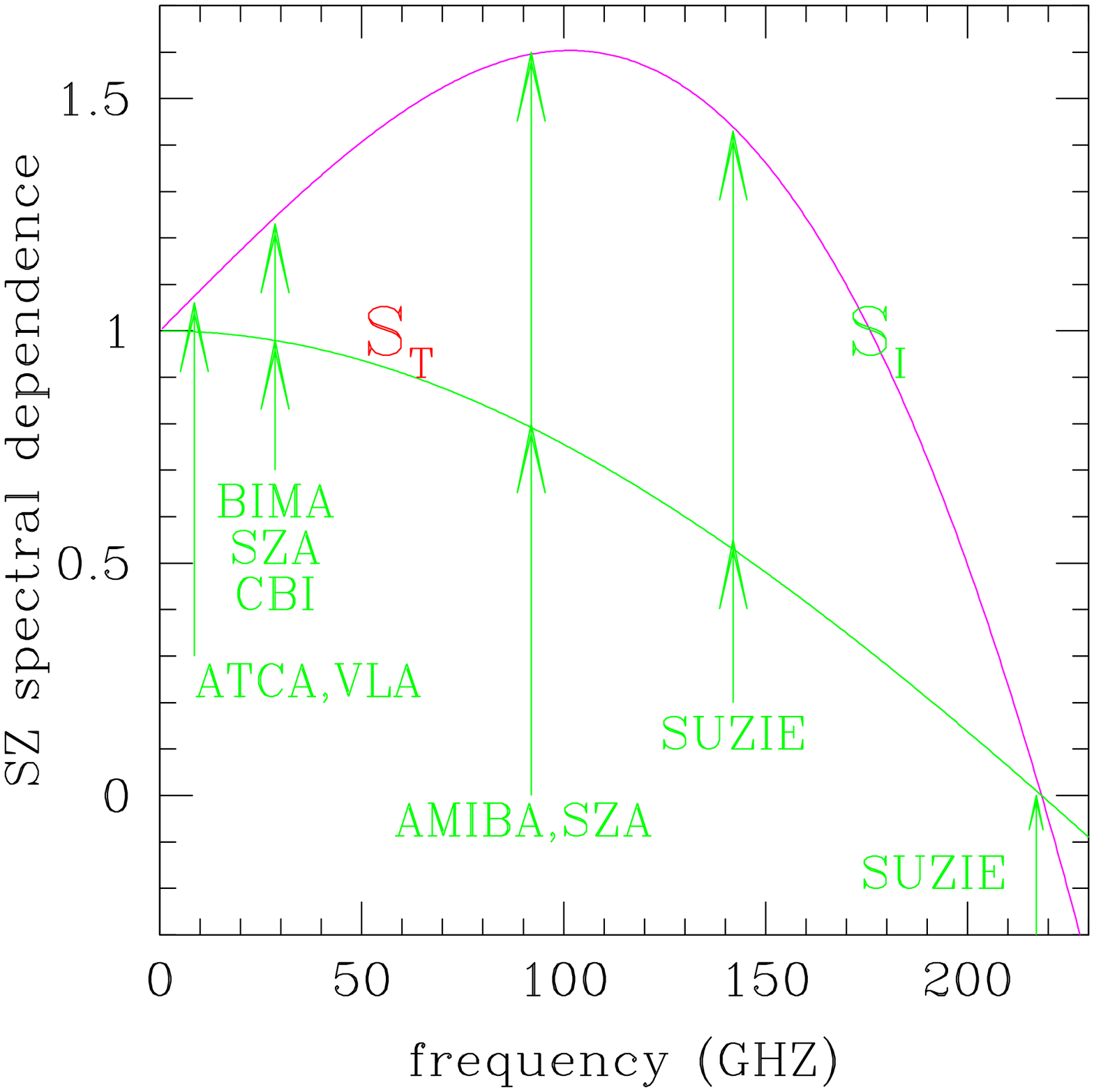}
\caption{
The SZ spectral dependence of CMB brightness fluctuation
$S_I(\nu)$ and temperature fluctuation $S_T(\nu)$. $\delta
I_{\nu}/I^{\rm CMB}_{\nu}=-2y S_I(\nu)$. $\Delta
T/T_{\rm CMB}=-2y S_T(\nu)$. We show the responses
of various experiments which are capable of detecting SZ signals such
as AMIBA: 90 Ghz \citep{AMIBA}, ATCA: 
$8.7$ Ghz \citep{Subrahmanyan00},  
BIMA: $28.5$ Ghz \citep{Dawson01}, CBI: $26$-$36$ Ghz, SUZIE: $142$
Ghz, $217$ Ghz, etc.  
\citep{Church97}, SZA: $26$-$36$ Ghz and $85$-$115$
Ghz and VLA: $8.4$ Ghz \citep{Partridge97}.
$S(\nu)\rightarrow 1$ when $\nu\ll 50$ Ghz. At $217$ Ghz, there is no
thermal SZ effect, enabling the measurement of other
CMB anisotropies such as the kinetic SZ effect. \label{fig:sx}
}
\end{figure}

The thermal SZ effect causes CMB temperature and intensity
fluctuations in the sky. At a given
position given by a unit vector $\hat{n}$ pointing from the earth to
some point on the sky,  the  fractional change in the CMB intensity $\delta
I_{\nu}/I_{\nu}$ depends on both the observing
frequency $\nu$ and the integral of the gas pressure along the line of
sight \citep{Zeldovich69}:
\begin{equation}
\frac{\delta I_{\nu}}{I_{\nu}}=-2y(\hat{n}) S_I [x(\nu)].
\end{equation}
For the CMB intensity, the spectral dependence for inverse compton
scattering off non-relativistic electrons is described by
$S_I(x)=\frac{xe^x}{e^{x}-1}\left(2-x/[2\tanh(x/2)]\right)$ where
$x\equiv h\nu/(k_B T_{\rm CMB})=\nu/57$ Ghz. The corresponding fractional
change in CMB temperature  $\Theta(\hat{n}) \equiv\frac{\Delta
T_{\rm CMB}(\hat{n})}{T_{\rm CMB}}$ is then:
\begin{equation}
\Theta(\hat{n},\nu)= -2y(\hat{n})
S_T[x(\nu)]=-2y\left(2-x/[2\tanh(x/2)]\right) . 
\end{equation}
In the Rayleigh-Jeans limit ($x\ll 1$), $S_I(x)=S_T(x)=1$.   
$S_I(x)$ reaches the highest response, $S(x)\simeq 1.6$
at $\nu \simeq 100$ Ghz, which is well matched to the AMIBA
frequency range of 80-100 Ghz. $S_T(x)$ monotonously decreases
with frequency. We
show $S_I(x)$  and $S_T(x)$ in fig. (\ref{fig:sx}) where the frequency
responses of various experiments which already have the CMB observation
data at small angular scales ($l>2000$) such as ATCA  \citep{Subrahmanyan00},  
BIMA \citep{Dawson01},
CBI\footnote{http://www.astro.caltech.edu/~tjp/CBI/}, SUZIE
\citep{Church97} and VLA 
\citep{Partridge97} are shown. 
The gas pressure dependence is described by the Compton $y$ parameter 
\begin{equation}
\label{eqn:y}
y(\hat{n})=\frac{\sigma_T}{m_e c^2} \int_0^{l_{\rm re}} n_e(l \hat{n})
k_BT_e(l\hat{n}) dl=
\frac{\sigma_T}{m_e c^2} \int_0^{l_{\rm re}} P_e(l\hat{n}) dl.
\end{equation}
$T_e$, $n_e$ and $P_e$ are the temperature, number density and
pressure of free electrons, respectively.  $dl$ is the proper distance interval
along the path of CMB photons.  $\sigma_T$, $m_e$ and $c$ have their
usual meanings as the Thomson cross section, electron mass
and the speed of light.  Since only free electrons participate in
Thomson scattering of the CMB photons, the integral is cut off at the
epoch of reionization $l_{\rm re}$.

Some important statistics of the SZ effect immediately come to mind.
The first order quantity is $\bar{y}$, the mean $y$ parameter averaged
over the whole sky, which measures the total thermal energy content of
the universe.  The angular variation in $y$ can be parameterized by the
two point correlation function of the temperature fluctuation
$\Theta$, or equivalently the angular power 
spectrum $C_l$.  For a Gaussian random field, these two parameters
would describe the statistics completely.  Since the SZ effect is
dominated by non-linear structures, non-Gaussianity may be
significant. So we investigate higher order statistics such as the
skewness and kurtosis of the $y$ parameter to quantify the
non-Gaussianity. 

The SZ effect contains contributions from all redshifts and it is
challenging to recover the smeared redshift information. We
have shown that cross correlating the SZ effect with a galaxy
photometric redshift survey, we can infer the redshift resolved IGM
pressure-galaxy cross correlation and the IGM pressure
auto correlation \citep{Zhang01a}.  This method is robust, but does
not capture all the information in  the SZ observation. In this paper, we
utilize the one point distribution function (PDF) of the $y$ parameter and
the distribution of peaks in $y$ to extract more information.

Since smoothing is always
present for a real experiment with a finite beam, we calculate the
statistics of the $y$ parameter  smoothed on a given angular scale
$\theta$.   If we are interested in
virialized objects, for 
example clusters and groups of galaxies, we would expect them to be
peaks in the smoothed or filtered $y$ maps. $N(y>y_p)$, the cumulative
distribution function (CDF) of  peaks with smoothed $y$ 
parameter bigger than certain value $y_p$,  is the raw  observable. 
The $y_p$ CDF  is the SZ analog to a luminosity function.
If we choose 
a top hat window so that the observation cone is large enough to
include an entire object such as a cluster and is small
enough that typically no  more than one such object  can be
found along each cone, then when the cone is centered at the center of
each object, a peak  $y\equiv y_p$ appears in the smoothed
map. This $y_p$ is directly related to the total gas mass and temperature of
individual object. Assuming a halo to be isothermal, the
total mass $M$ of a halo is related to the gas temperature $T$ by
$M/M_8=(T/T_8)^{3/2}$. $M_8=1.8\times 10^{14} (\Omega_0/0.3) h^{-1} {\rm M}_{\sun}$ is the 
mass contained in a $8 {\rm  h}^{-1}{\rm Mpc}$ sphere of the universe
of mean density today, which is
roughly the mass scale of clusters.  $T_8(z)$ is the corresponding
temperature of a halo with mass $M_8$ at redshift $z$. We then obtain 
\begin{equation}
\label{eqn:ypeak}
y_p=
\frac{M_8f_g}{\mu m_H} \frac{\sigma_T}{d_A^2 \Delta \Omega}
\frac{k_B T_8}{m_e c^2} \left( \frac{M}{M_8} \right)^{5/3}.
\end{equation}
Here, $\Delta \Omega$ is the solid angle of the cone, $f_g$ is the
gas fraction of halos and $d_A$ is the angular diameter distance. 
For clusters and groups, the typical angular size at $z=1$
is about $1^{'}$. For the present cluster number density $n(T>2 {\rm keV})\sim 
10^{-5} h^{-3} {\rm Mpc}^3$ \citep{Pen98b}, the average number of clusters in a
cone with angular radius   $\theta \sim 20^{'}$ projected to $z\sim 2$
is about one allowing for the evolution of cluster number density. So,
the size of the smoothing scale for the peak 
analysis should be between these two scales. In this case, the
$N(y>y_p)$ is just the number of halos with $y>y_p$, which is given by: 
\begin{equation}
\label{eqn:cdf}
N(y>y_p)=\int_0^{r_{{\rm re}}} \chi^2 dr/\sqrt{1-{\rm K}r^2} \int^{\infty}_{M(y_p,z)} \frac{{\rm
d}n}{{\rm d}M} {\rm d} M. 
\end{equation}
$\frac{{\rm d} n}{{\rm d}M}(M,z)$ is the halo comoving number density
distribution  function and is well described by the Press-Schechter formalism
\citep{Press74}. $\chi$, $r$ and ${\rm K}$ are the comoving distance, radial
coordinate and curvature of the universe, respectively. The subscript 're'
means the reionization epoch. $M(y_p,z)$ is the mass of the halo with
smoothed $y=y_p$ given by eqn. (\ref{eqn:ypeak}). Eqn. (\ref{eqn:cdf})
has two applications. Firstly, given a SZ survey and the best
constrained cosmological 
parameters as determined  by CMB experiments, Type ${\rm I}_a$ SN,
weak lensing, etc. and a SZ survey,  the only unknown  
variable in eqn. (\ref{eqn:cdf}) is $T_8(z)f_g(z)$, which then is
uniquely determined.  $T_8$ is robustly predicted since it is
mainly determined by ${\rm M}_8$  through hydrostatic
equilibrium and has only weak dependence on the thermal
history.  For example, comparing
the cluster temperature function as inferred from simulations with the
Press-Schechter formalism, \citet{Pen98b} found that
$T_8=4.9 (1+z) \Omega_0^{2/3}  \Omega(z)^{0.283} {\rm keV}$ for a
$\Lambda$CDM universe. The above relation is sufficient to extract
the evolution 
of the gas fraction $f_g$, which is very sensitive to the thermal
history. For example, non-gravitational energy injection decreases
$f_g$. Secondly, the number of clusters strongly depends on
cosmology, as characterized by $\frac{{\rm d}n}{{\rm d}M}$. Given a good understanding of cluster temperature
and gas fraction, eqn. (\ref{eqn:cdf}) allows one to constrain
cosmological parameters. In combination with the measurement of
cluster redshifts, 
this method is more sensitive \citep{Weller01}.  In \S \ref{sec:AMIBA},
we will study the 
statistics of these peaks 
for maps filtered in optimal ways to measure clusters of galaxies.

    With the measurement of the cluster SZ temperature distortion and
follow up X-ray observations of cluster X-ray flux $F_X$ and
X-ray  temperature $T$,  cosmological parameters can be
constrained. For a cluster with electron number   
density $n_e$, proper size $L$ and angular size $\theta$,  $F_X\propto
n_e^2 \Lambda(T) L^3/d_L^2$ and $y\propto n_e T L$. Here, $\Lambda(T)$ is the
X-ray emissivity temperature dependence. Then the
luminosity distance $d_L  (z) \propto
\frac{y^2}{F_X} \frac{\Lambda(T)}{T^2} \frac{\theta}{(1+z)^2}$.  The only
uncertainties in this relation are the  intracluster gas profile and
metalicity which affects the X-ray emissivity. These properties could
be modelled and are potentially observable. In this 
sense, the SZ effect can be used as a cosmological
distance indicator \citep{Silk78,Barbosa96,Mason01b,Fox01}.  Since dark
energy dominates at low redshift where the SZ observation and
X-ray observation of clusters are relatively straightforward, SZ
clusters are a potential probe to constrain the equation of state for
the dark energy.

\section{Simulations}
\label{sec:simulation}

In this section, we describe our SZ simulation used to investigate the
above statistics. We will also use our simulation to provide SZ maps
for our estimation of the sensitivity of upcoming SZ experiments (\S
\ref{sec:AMIBA}) and its effect on data analysis strategies.  We
used a moving mesh hydrodynamics (MMH) code \citep{Pen98a}. It
features a full curvilinear total-variation-diminishing (TVD) hydro
code with a curvilinear particle mesh (PM) 
N-body code on a moving coordinate system.  The full Euler equations
are solved in an explicit flux-conservative form using a second order
TVD scheme.  The curvilinear
coordinates used in the code are derived from a gradient of the
Cartesian coordinate system.  If $x^i$ are the Cartesian coordinates,
the curvilinear coordinates are $\xi^i = x^i + \partial_\xi^i\phi({\bf
\xi})$.  The transformation is completely specified by the single
potential field $\phi({\bf \xi},t)$.  The potential deformation
maintains a very regular grid structure in high density regions.  The
gravity and grid deformation equations are solved using a hierarchical
multigrid algorithm for linear elliptic equations.  These are solved
in linear time, and are asymptotically faster than the FFT gravity
solver.  At the same time, adaptive dynamic resolution is achieved.
During the evolution any one constraint can be satisfied by the grid.
In our case, we follow the mass field such that the mass per unit grid
cell remains approximately constant.  This gives all the dynamic range
advantages of smooth particle hydro (SPH) combined with the speed and high resolution of grid
algorithms.  The explicit time integration limits the time step by the
Courant condition.  To achieve a reasonable run time, we limit the
compression factor to a factor of 5 in length, corresponding to a
factor of 125 in density.  Most SZ contributions arise below such
densities, giving a diminishing return to go to higher compression
factors.

\begin{figure}
\plotone{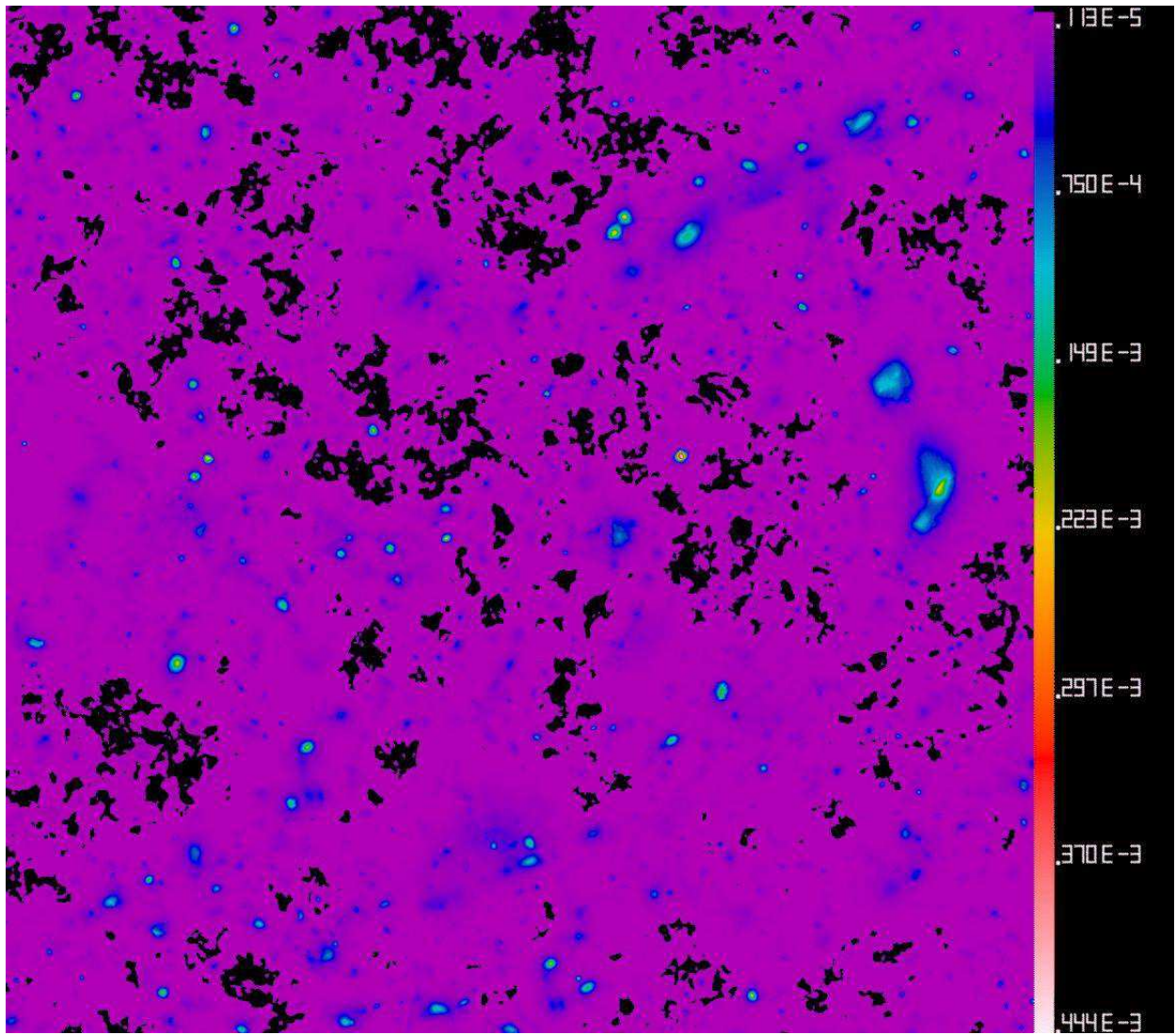}
\caption{One typical thermal SZ map in our moving mesh hydro (MMH)
code simulation. The cosmology
is a $\Lambda$CDM with $\Omega_0=0.37$, $\Omega_{\Lambda}=0.63$,
$\Omega_B=0.05$, $h=0.7$ and $\sigma_8=1.0$. The map size is
$1.19^{\circ} \times 
1.19^{\circ}$. The color represents the SZ temperature fluctuation in the
Rayleigh-Jeans regime $\Delta T/T=-2 y$. We have omitted the negative
sign. The SZ map clearly shows the
structures with angular scale $\sim 1{'}$, which is the typical scale
of clusters and groups. \label{fig:tszmap}}
\end{figure}

\begin{figure}
\plotone{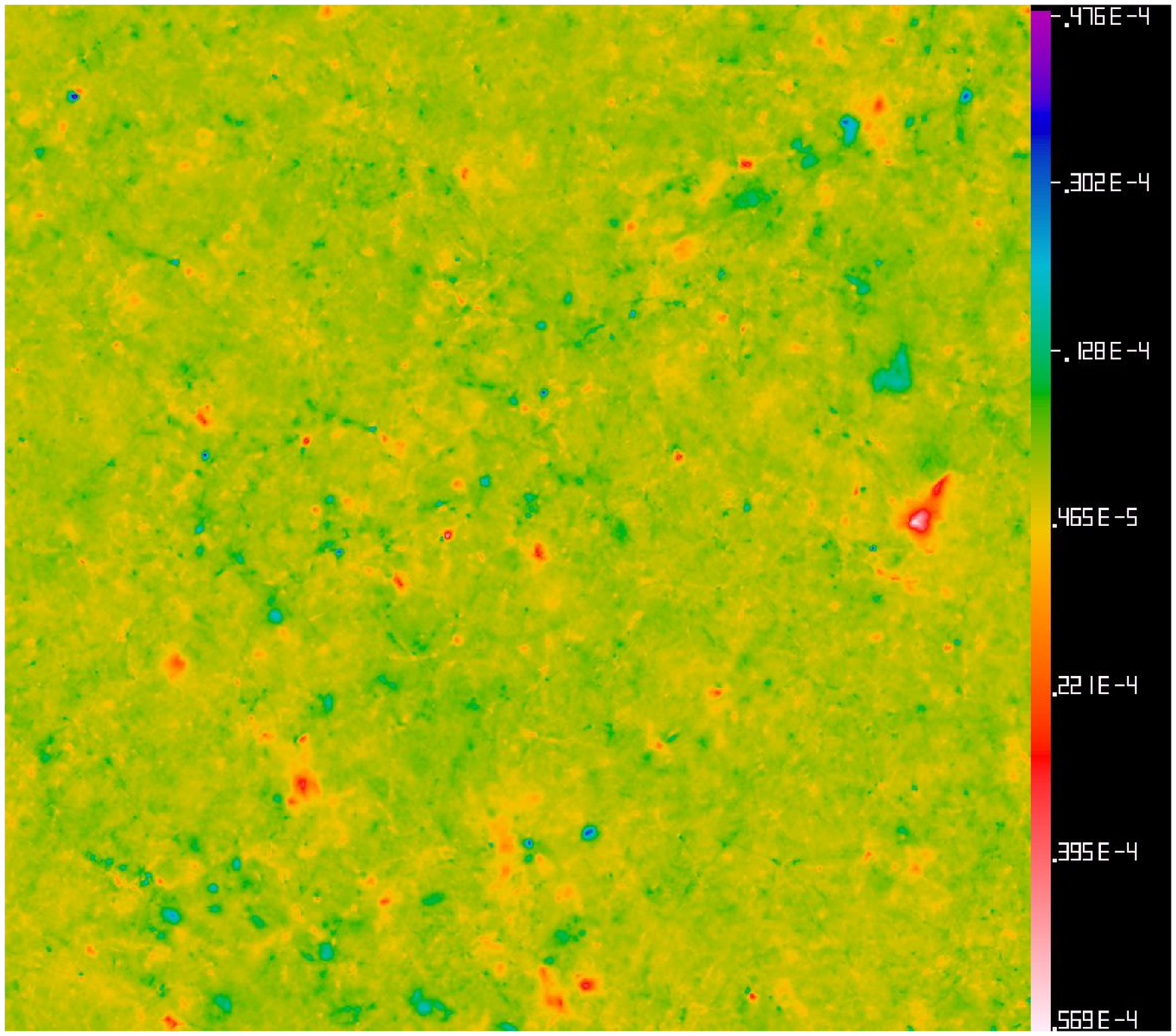}
\caption{The kinetic SZ map in the same simulation at the same
field of view as fig. \ref{fig:tszmap}.  The amplitude of temperature
fluctuation of the kinetic SZ 
effect is generally one order of magnitude lower than the thermal SZ
effect.  Gas in the positive $\Delta T/T$ regions is approaching us and
those in the negative $\Delta T/T$ regions is moving away. \label{fig:kszmap}}
\end{figure}

The parameters we adopted in our $512^3$ simulation are
$\Omega_0=0.37$, $\Omega_{\Lambda}=0.63$, $\Omega_B=0.05$, $h=0.7$,
$\sigma_8=1.0$, power spectrum index $n=1$, box size $L=100 h^{-1}$
Mpc and smallest grid spacing $40 h^{-1}$ kpc.  
The simulation used $30$ GB memory and took about three weeks
($\sim 1500$ steps) on a 32 processor shared memory Alpha GS320 at CITA
using Open MP parallelization directives. 
During the simulation we
store 2D projections through the 3D box at every light crossing time
through the box. The projections are made alternatively in the x, y, z
directions to minimize the repetition of the same structures in the
projection. We store projections of thermal SZ, kinetic SZ, gas and
dark matter densities. For the thermal SZ, we store $2 \Delta
y=\frac{2 \sigma_T}{m_ec^2} P_e L$ as given by eqn. (\ref{eqn:y}).  Our
2D maps are stored on $2048^2$ grids. As tested by \citet{Seljak01},
this preserves all the information down to the finest grid
spacing. After the simulation, we stack the SZ sectional maps
separated by the width of simulation box, randomly choosing the center
of each section and randomly rotating and flipping each section.  The
periodic boundary condition guarantees that there are no
discontinuities in any of the maps. We then add these sections onto a
map of constant angular size.  Using different random seeds for the
alignments and rotations, we make $40$ maps of width $1.19$ degrees to
calculate the SZ statistics. The mean $y$ parameter in these $40$ maps
$\bar{y}\simeq4 \times  10^{-6}$.  The mean $y$  parameter is
still below the upper limit $1.5\times 10^{-5}$ from COBE FIRAS
\citep{Fixsen96}. One  typical thermal SZ sky map and a kinetic SZ
sky map at the same field of view are shown in
fig. \ref{fig:tszmap} and fig. \ref{fig:kszmap}, respectively. With
these maps, we calculate the SZ 
power spectrum, the SZ non-Gaussianity, the $y$ PDF and $y_p$ CDF.

\subsection{The SZ power spectrum}
\begin{figure}
\plotone{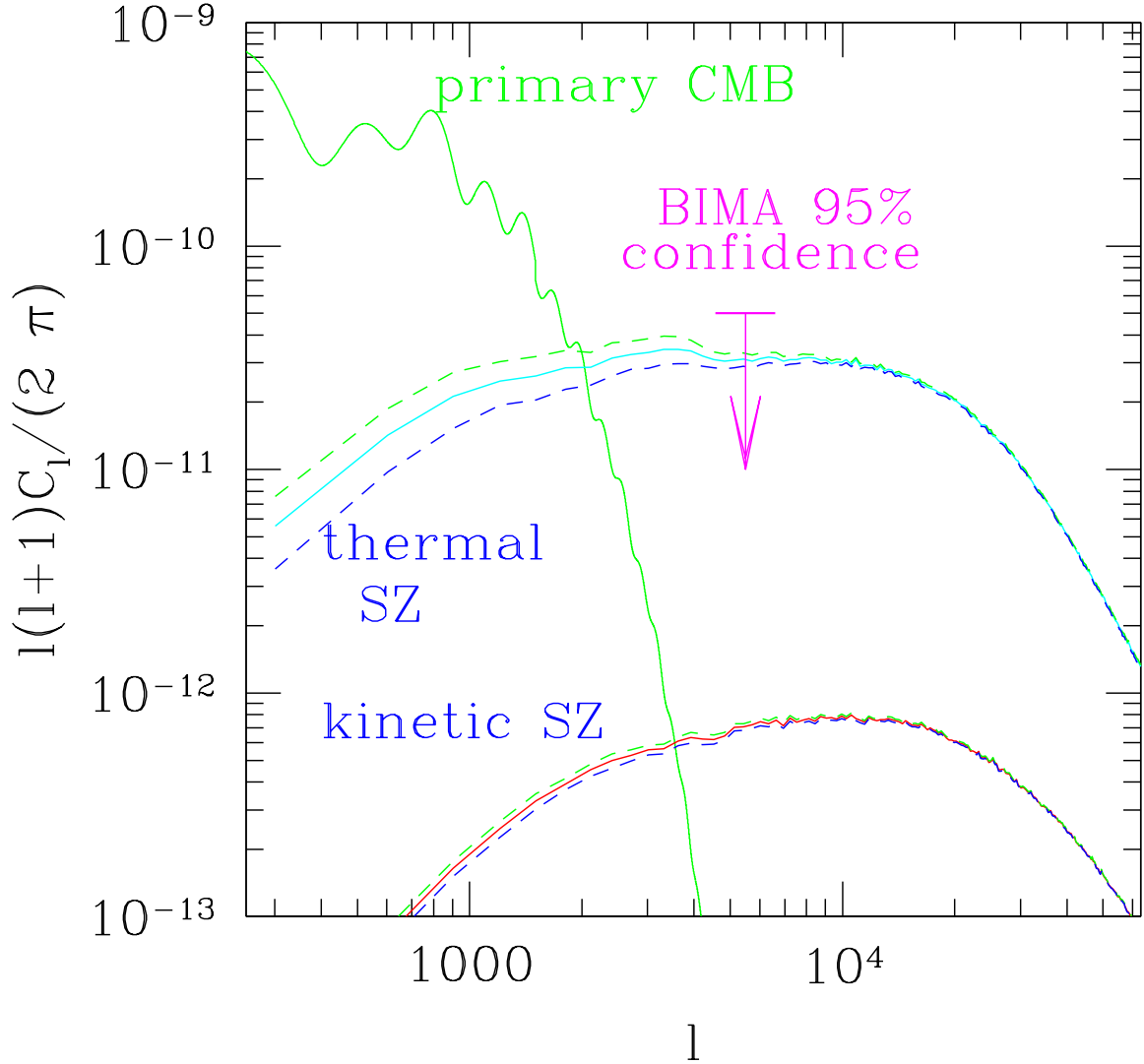}
\caption{The thermal and kinetic SZ effect power spectra in our
simulation.They are averaged over $40$ maps. Dash lines are the
corresponding $1\sigma$ upper limit and lower limit  of the mean
power spectrum, respectively.  Both effects peak at
$l\sim 10^4$. For the 
thermal SZ effect, the power spectrum is almost flat in the range of
$2000\lesssim l \lesssim 15000$. As a comparison, we show the BIMA
result \citep{Dawson01} under Gaussian assumption. The comparable
amplitude between theory and 
observation puts strict constrain on the understanding of the SZ effect. 
\label{fig:cl}}
\end{figure}

The power spectrum in the
Rayleigh-Jeans regime averaged over 40 maps is shown in
fig. \ref{fig:cl}.  The thermal SZ 
power spectrum is fairly flat, with a broad peak at $l\sim 2000$-$10^4$
and  a typical fluctuation amplitude $\Theta \sim 5.5 \times10^{-6}$.
It begins to dominate over the primary CMB at 
$l\sim 2000$. The shape of the kinetic SZ power spectrum is
similar to the  thermal one but the amplitude is  about $30$
times smaller.

Comparing to available CMB observations, our thermal SZ power spectrum
is well below the upper  limit ($95\%$ confidence) of ATCA
\citep{Subrahmanyan00}, SUZIE \citep{Church97} and VLA
\citep{Partridge97}.  $C_l$ at $l\sim 2000$ is consistent with recent
indications  from the CBI experiment  
\citep{Mason01a,Sievers01} and may suggests a high $\sigma_8\simeq
1.0$. But our result is higher than the BIMA $1$-$\sigma$ result
thought it is consistent with the upper limit ($95\%$ confidence) of the
BIMA result\citep{Dawson01}.  We will further discuss these issues
below.

\begin{figure}
\plotone{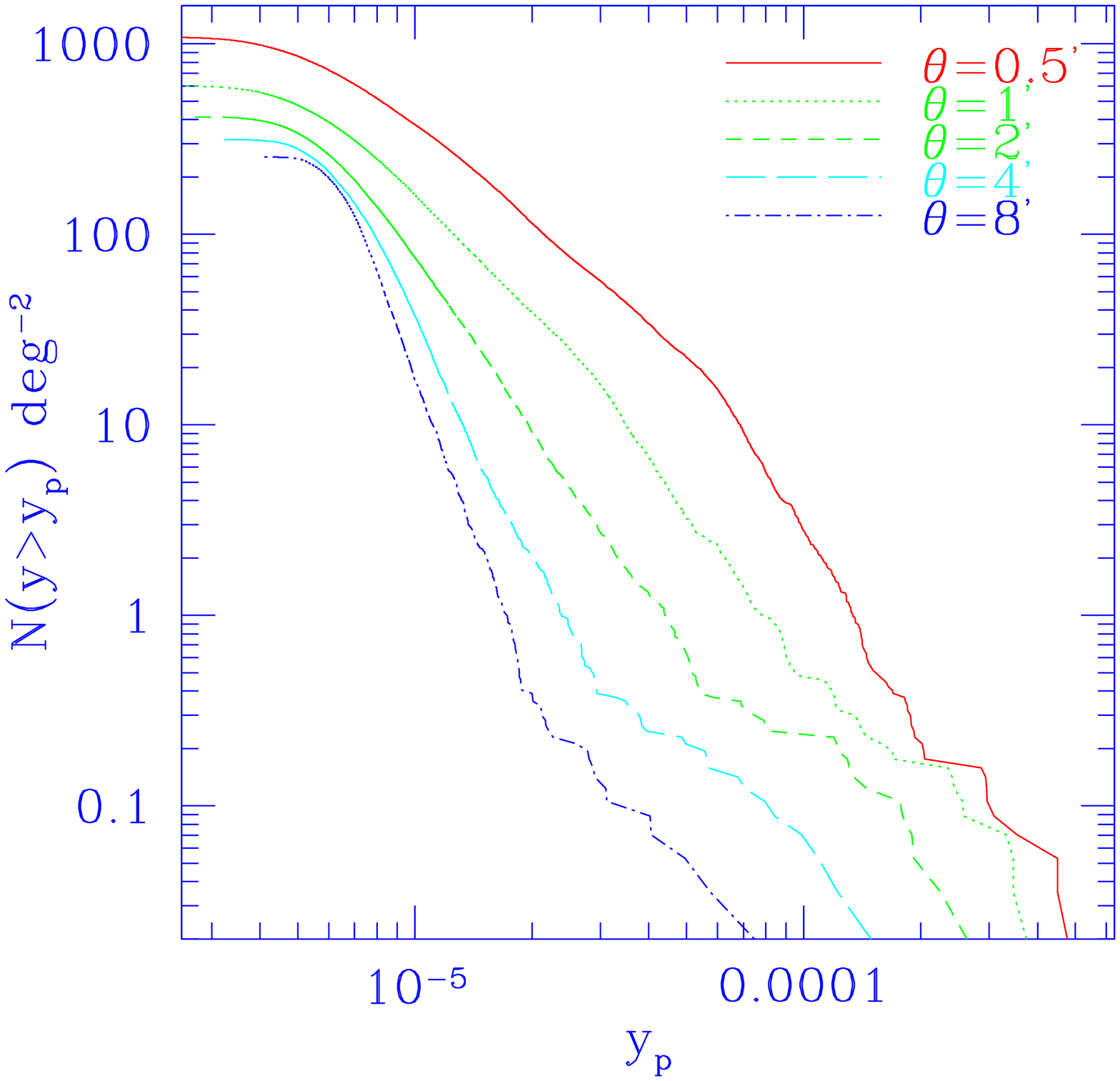}
\caption{$N(y>y_p)\equiv \int^{\infty}_{y_p} \rho(y) dy$ is the number
of peaks with Compton parameter $y$ bigger than $y_p$. The result is
averaged over $40$ maps. The smoothing function we adopt is the top
hat window function with radius $\theta$. $N(y>y_p)$ as a function of
$\theta$ is a direct measurement of number of structures with
scale larger than $\theta$. The quick drop of $N_{y>y_p}$ from
$\theta=0.5^{'}$ to $\theta=1^{'}$ shows that there are a lot of
arcminute scale structures. We believe that these structures with
large $y$ is responsible for the divergence of SZ simulations towards
high resolution.  \label{fig:peak}} 
\end{figure}

 Published theoretical predictions differ a lot in both amplitude and shape
(see \citet{Springel01} for a review). Our power
spectrum has a higher amplitude than all previous predictions. It is also
significantly flatter at the range $2000<l<15000$. The difference in
amplitude  can be  explained by the strong dependence of the SZ effect
on cosmological parameters, 
especially $\sigma_8$. One expects $y\propto
\Omega_B h \sigma_8^{2--3}$ and $C_l \propto (\Omega_B h)^2
\sigma_8^{6--9}$ as predicted by various authors
\citep{Komatsu99,Seljak01,Zhang01a}. For example, \citet{Seljak01} used
the same MMH code with lower resolution ($256^3$) to simulate a
universe with $\sigma_8=0.8$.  This $\sigma_8$ difference accounts for
a factor of $1.6\sim 2.0$ difference in the mean $\bar{y}$ estimation
and $3.8 \sim 7.5$ difference in the $C_l$ estimation.  After
accounting for these effects, our power spectrum is consistent with
theirs at small angular scales ($l>3000$). But the difference in shape
can not be explained in this way. For example, even after
accounting for the effect of cosmological parameters, our power
spectrum is still much larger at large angular scales
($l\sim 1000$) than that of \citet{Seljak01} (at $l\sim 1000$,
$\sim 2$ times larger). This flatness behavior may be a
manifestation of the resolution effect.

We notice that there are numerous high $y$ regions
of arcminute or sub-arcminute scales in SZ maps as seen in
fig. \ref{fig:tszmap}. To quantify this phenomenon, we show the dependence of
$N_{\theta}(y>y_p)$ on smoothing angular size $\theta$
(fig. \ref{fig:peak}). $N_{\theta}(y>y_p)$ is a measure of
number of structures with angular size larger than or 
comparable to $\theta$. Fig. \ref{fig:peak} shows that when the
smoothing scale increases from $0.5^{'}$ to $1^{'}$, $N(y>y_p)$ drops
significantly in high $y$ regions. This behavior suggests the existence
of numerous sub-arcminute, high $y$ 
structures in SZ maps. Higher resolution reveals more such
structures. Increasing in the number of these objects
increases the amplitude of the power 
spectrum while making the power spectrum flatter around the peak.  The
first effect is obvious and the second one  can be 
explained by the Press-Schechter picture. The SZ power spectrum is
dominated by the halo gas pressure profile $f_P(r)$ at all interesting
angular scales  \citep{Komatsu99}.  For a singular isothermal sphere (SIS), 
$f_p(r)=n_e(r)k_BT(r) \propto n_e(r)\propto r^{-2}$. Its projection
along the line of sight then has a $\theta^{-1}$ radial dependence,
which produces an 
angular auto correlation function of shape $\ln(\theta)$. So the
resulting power spectrum is 
nearly flat. A more accurate way to see the flatness behavior is to
adopt the Limber's equation.  The 3D gas pressure
power spectrum $P_p(k)\propto \delta^2_p(k)\propto k^{-2}$, where
$\delta_p(k)$ is the Fourier transform of the gas pressure profile
$f_p(r)$.  From Limber's equation,
$C_l\propto \int P_p(l/\chi(z),z) f(z) d\chi \propto l^{-2}$, thus
$l(l+1)C_l/(2\pi)\propto l^0$.  Here, $f(z)$ is the redshift
dependence of the SZ effect.  The halo mass function also plays a role for
the flat power spectrum.    
Since SIS profile does not apply to the
core of halos,  a given cluster is no longer SIS at scales smaller
than its core size. Smaller clusters take over and extend the flat
power spectrum.  The power spectrum in our simulation
clearly shows this flatness and suggests the role of  these
sub-arcminute halos. 
These halos also explain the discrepancy between our simulation and
analytical predictions.  $1^{'}$ corresponds to the comoving size $\lesssim 0.8
h^{-1}$ Mpc for $z<1$ where the dominant SZ contribution
comes from.  This physical size corresponds to groups of
galaxies. Current predictions from the Press-Schechter picture assume
a lower mass limit cutoff corresponding to the mass scale of
groups. In the hierarchical method, the gas window
function is a free parameter, which has an implicit dependence on this lower mass cutoff (they
can be related by the gas density dispersion). The  cut off for
contributions from groups of galaxies in analytical models results in
a smaller $\bar{y}$ and power spectrum.

If only gravitational heating is included, high mass halos and low mass
halos should have comparable gas fractions.  The Press-Schechter
formalism predicts many more halos towards the low mass end, so one
expects pure gravity simulations to have increasing power  spectra
with increasing high resolution.  Non-gravitational 
heating avoids such a divergence. As described in the halo model of
\citet{Pen99}, non-gravitational heating has two effects. Firstly,
the energy injection makes the halo gas less clumpy. Then the
contribution to  the power spectrum at smaller angular scales
decreases relative to larger angular scales. This could explain the
slightly differences between our simulation, CBI and BIMA
results. Secondly, the non-gravitational energy injection increases 
the thermal energy of the gas. For halos with
mass lower than some threshold, the gravity can not hold gas and
most of the gas is ejected from these halos, as must have been the case for
galactic size halos.  This provide a reasonable lower mass limit cut
off in the Press-Schechter picture. Thus the amplitude and shape of
the SZ power spectrum is a sensitive measurement of the
non-gravitational energy injection. To obtain a better understanding
of the SZ effect, other physicical processes such as radiative cooling need
to be considered. Since radiative cooling through thermal
bremsstrahlung is a $\rho^2$ process, it becomes relevant only at
scales $\lesssim 100$ kpc, which corresponds to $l\gg 10^4$.  These
angular scales are not observable by any planed experiments, so we
neglect the discussion of  radiative cooling in this paper. Our 
current simulation has reached the resolution needed to see the
contribution from small halos and the predicted SZ amplitude is
already near the observed values. We expected to be able to
observe the effects of non-gravitational heating from galaxy winds and
other sources in the upcoming experiments.

In order to solve the discrepancy problem completely, differences
between codes must be considered. Our
group is currently running different codes with identical initial
condition, identical cosmological parameters and 
various resolutions from $64^3$ to $512^3$ \citep{Codes}.

Our prediction is consistent with the BIMA $95\%$ confidence  result.
But the comparison between observations and theoretical predictions
needs further investigation. On the simulation part, as we discuss
above, the detailed normalization depends on $\sigma_8$,
non-gravitational heating and resolution effects.  Observations will put 
strong constrains on these aspects. On the observational part, the BIMA
result needs to be reconsidered. It
covers the sky where no strong SZ temperature
distortion due to known galaxy clusters exists. Since clusters
contribute a significant if not dominant fraction to the SZ power
spectrum, the BIMA  result may be smaller than the
statistical mean value. Furthermore, the  error of the BIMA result is
estimated under the Gaussian
assumption, but as we will see below, the SZ effect is highly
non-Gaussian around the BIMA central multipole $l=5530$. The strong
non-Gaussianity increases the intrinsic error  of the SZ power spectrum
measurement. According to fig. \ref{fig:ngs}, the error in the power
spectrum measurement caused by
the SZ effect is about $3$ times larger than the corresponding
Gaussian case.

Nontheless, these results demonstrate the convergence between theory
and observation.  In the near future, routine and accurate measurement
of the SZ effect 
will be possible in random fields. It will  put stronger requirement
for our theoretical understanding of the SZ effect and allows us to
study the effects of non-gravitational heating, radiative cooling and
the thermal history of the IGM.

\subsection{The SZ non-Gaussianity}
\begin{figure}
\plottwo{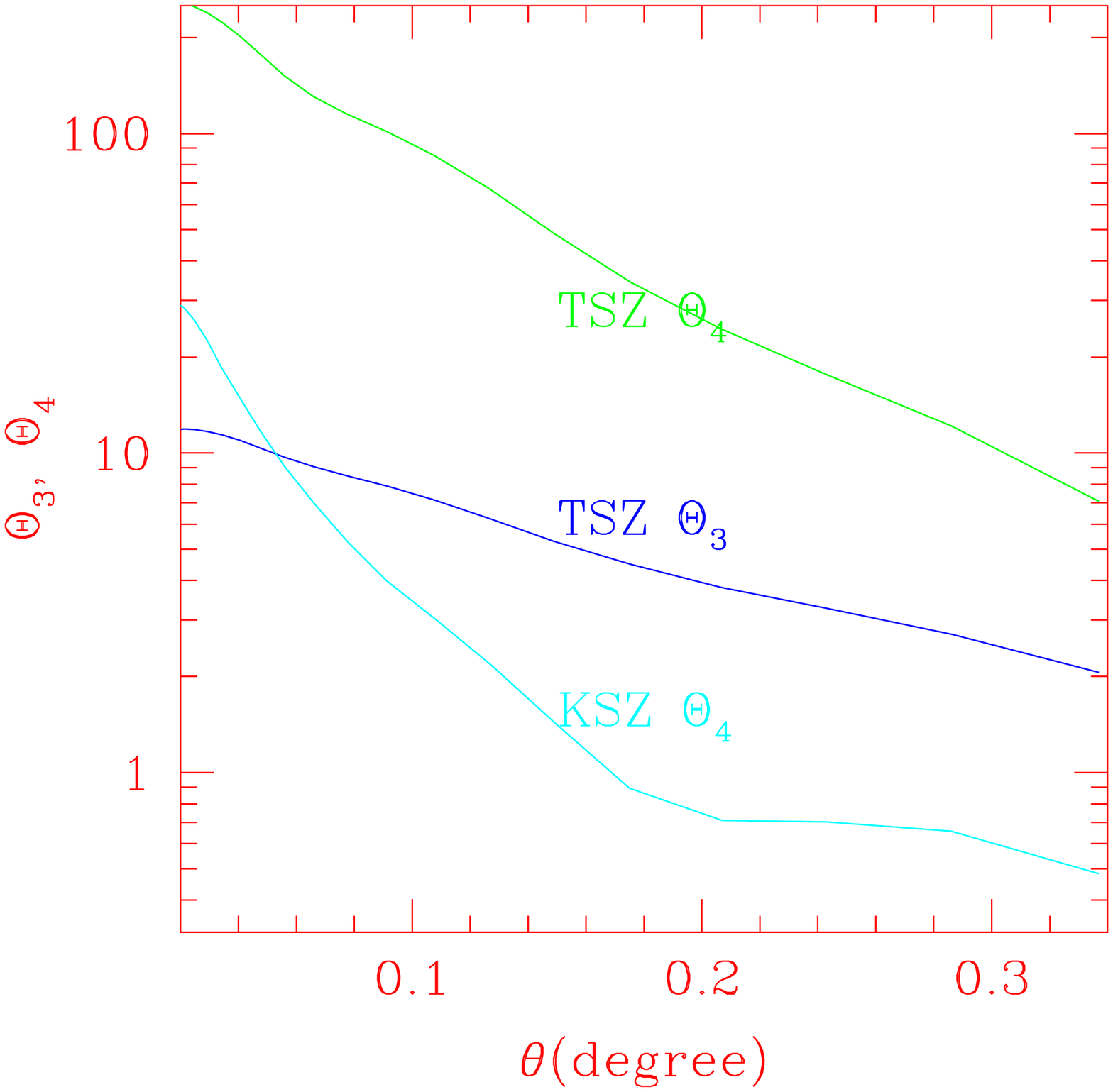}{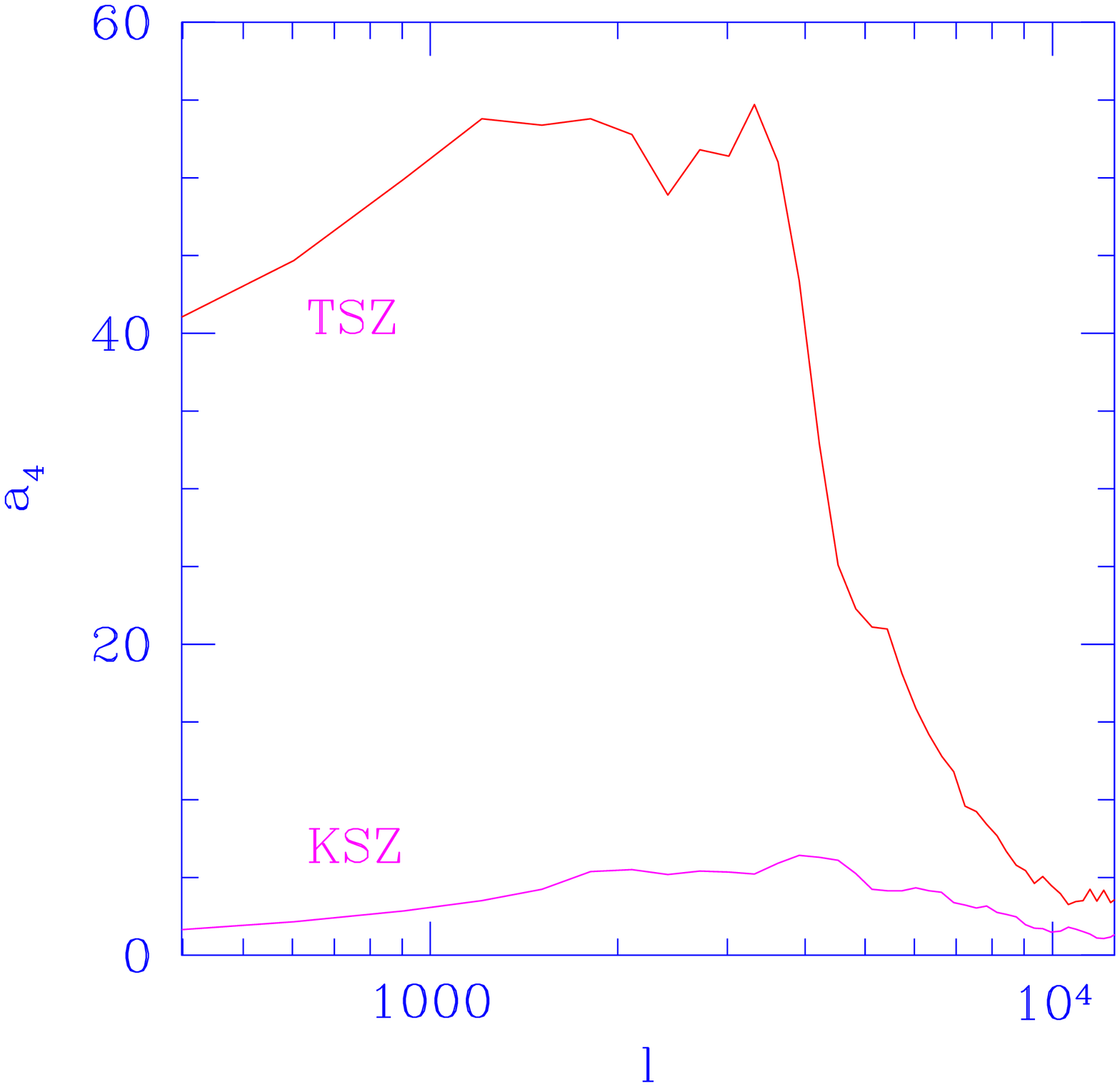}
\caption{The non-Gaussianity of the SZ effects in real space (left
panel) and multipole space (right panel). In real space, we measure
the skewness
$\Theta_3$ and kurtosis $\Theta_4$. For symmetrically distributed signals,
$\Theta_3=0$, as in the kinetic SZ case. For a Gaussian signal,
$\Theta_4=0$. The left panel shows that the thermal SZ effect 
is highly non-Gaussian even to the angular size $\theta\sim
0.3^{\circ}$. The kinetic SZ is mildly non-Gaussian and approaches  
Gaussian soon towards large angular scales. 
The right panel shows the
non-Gaussianity in multipole space. $a_4$ is the kurtosis in the
multipole space. Because multiple
modes are not local and 
averaged over many uncorrelated patches, the corresponding 
non-Gaussianity is much smaller than in the real
space. \label{fig:ngs}}
\end{figure}

In contrast to the primary CMB, the SZ effect is non-Gaussian, arising
from the non-linearity of the intervening gas.  This non-Gaussianity
affects the error analysis and may help
to separate  the SZ effect from the primary CMB in 
observation. To quantify these effects, we smooth SZ maps
using a top hat window of radius $\theta$ and measure the kurtosis of
the smoothed $y$. The kurtosis
$\Theta_4\equiv \frac{\langle y-\bar{y} \rangle 
^4}{\sigma_y^4}-3$ is generally a function of 
smoothing scale. For a
Gaussian signal, 
$\Theta_4=0$. Fig. \ref{fig:ngs} shows that $\Theta_4\simeq 200$ at
small scales as $\theta\rightarrow
0$. This result is 
consistent with the prediction from our hierarchical model approach of
the SZ effect \citep{Zhang01a}. We predicted, at small angular scales,
$\Theta_4\sim
S_4\sigma_g^2(z\sim 1) \sim 200$. Here, $S_4\equiv\frac{\langle\delta^4\rangle}{\langle \delta^2 
\rangle^3}\sim 40$ is a hierarchical model coefficient
\citep{Scoccimarro99} and $\sigma_g(z\sim1)\sim 5$ is the gas density
dispersion at $z\sim1$.  At large angular scales $\theta\sim 20^{'}$, 
$\Theta_4\gg 1$ and reflects the strong non-Gaussianity at this
scale. It means that, at this angular scale, the dominant contribution
to the $y$ 
parameter is from highly non-linear regions and there is still strong
correlation at angular scales down to  $l\sim 1000$, as can be seen from
the SZ power spectrum.  

\begin{figure}
\plotone{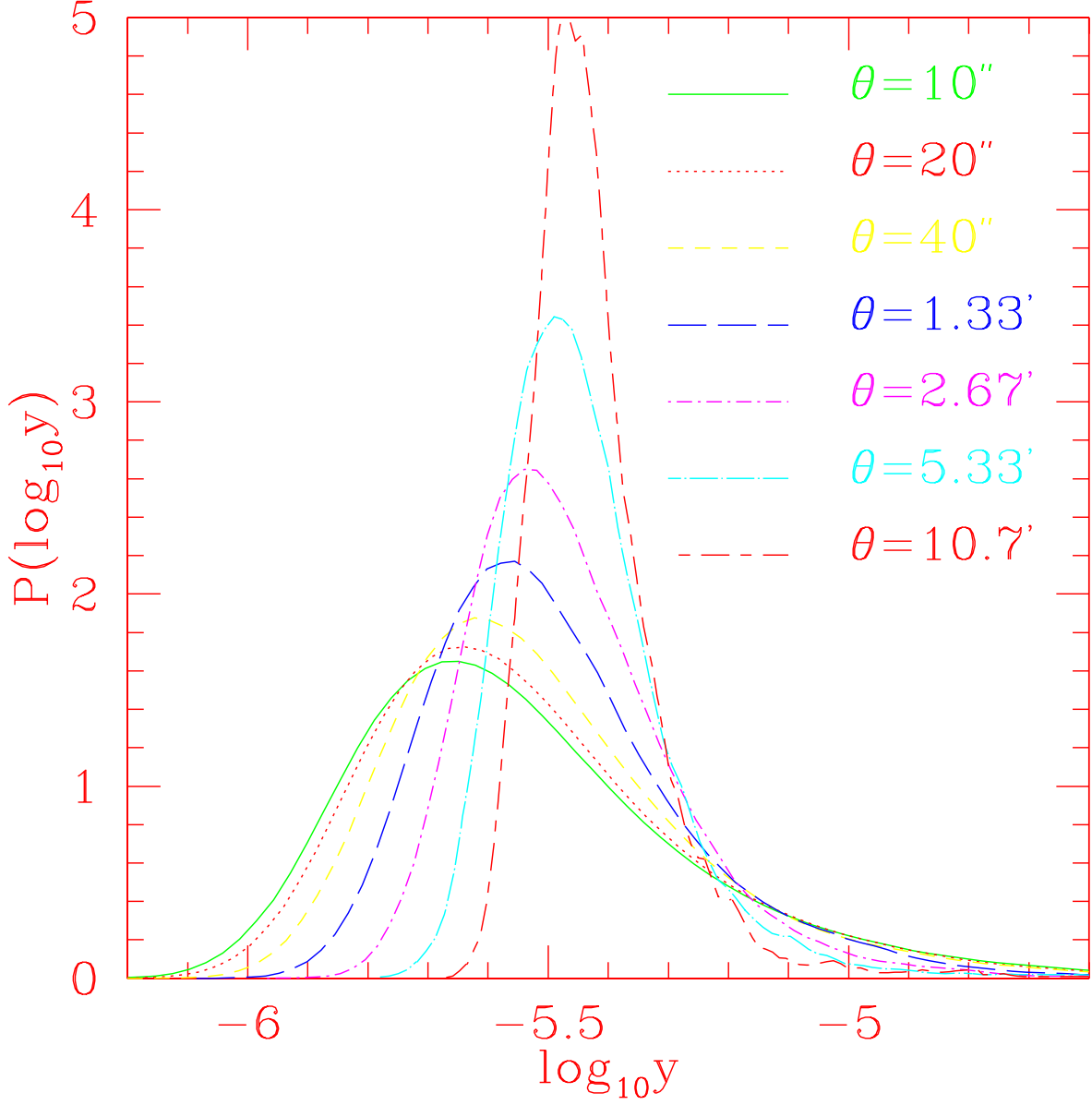}
\caption{The $\log_{10}{y}$ PDF $P(\log_{10}{y})$. We use the top hat
window with radius $\theta$ to smooth the SZ map.  $P(\log_{10}{y})$ is roughly
Gaussian. So the $y$ PDF $P(y)=P(\log_{10}{y})/(y \ln{10})$ is
non-Gaussian even with a large smoothing scale as shown in
fig. \ref{fig:ngs}. \label{fig:ydis}} 
\end{figure}

The kurtosis in multipole space $a_4\equiv\langle
|a|^4 \rangle/\langle |a|^2 \rangle^2-3$ . As usual, $a\equiv
a_{lm}$ defined by $\Theta(\hat{n})\equiv \sum_{lm} a_{lm}Y_l^m(\hat{n})$ are
multipole modes of $\Theta$ and $C_l\equiv
\sum_{-l}^{l} |a_{lm}|^2/(2l+1)\equiv \langle |a|^2\rangle$. Here,
$Y_{l}^m$ are the  spherical harmonics. $\langle \ldots \rangle$ is
averaged over all $m=-l,\ldots,l$ and all maps. We calculate $a_4$
indirectly through the map-map
variance $\sigma_M(l)$ of $C_l$, which is related to $a_4$ by $\sigma_M(l)=C_l
\sqrt{(a_4+2)/[(2l+1)\Delta l f_{\rm map}]}$.   $f_{\rm map}$ is the 
fractional sky coverage of each map which reflects the cosmic variance
and $\Delta l=2\pi/\theta_{\rm map}$ is the $l$ bin size in our grid
map with $\Theta_{\rm map}=1.19^{\circ}$. The factor $2l+1$ arises because
$C_l$ is averaged over $2l+1$ independent $a_{lm}$ modes.  
In our calculation, $C_l$ of each map is obtained using FFT under the flat sky
approximation. Then we obtain the map-map variance
$\sigma_M(l)$. 
One  might expect $a_4$ to have similar behavior as $\Theta_4$ at corresponding
scales, for example, at small angular scales (large $l$), $a_4\gg
2$. But fig. \ref{fig:ngs} shows the opposite. $a_4\sim 50$
around $l\sim 1000$, rises slowly until $l\sim 3000$  and
approaches Gaussian ($a_4=0$) quickly after that. 
In fact,  $\Theta_4$ and $a_4$ cannot be directly
compared.  The multipole modes are non-local, and averaged over many
patches.  If these patches are spatially uncorrelated, then the central
limit theorem assures the multipole modes to be Gaussian independent of
the actual non-Gaussianity of the patches. From this viewpoint,
we can estimate the relation between $a_4$ and $\Theta_4$. 
$Y_l^m(\hat{n})$ are quickly fluctuating functions with period $\Delta
\theta\simeq 
2\pi/l$. Two signals separated by a distance larger than $4\Delta
\theta$ will have effectively no correlation. So, for a $l$ mode,
$a_{lm}$ is approximately the sum of $N\sim \theta^2_{\rm map}/(4\Delta
\theta)^2$ uncorrelated patches. The non-Gaussianity of each patch is
roughly characterized by $\Theta_4$. Then, $a_4\sim
\Theta_4/N$, which may explain the behavior of $a_4$.

The shape of $a_4(l)$ is similar to the shape of the pressure bias
$b_p(k,z)$ (fig. 2, \citet{Zhang01a})
defined by $b^2_p(k,z)\equiv P_P(k,z)/P_{\delta}(k,z)$ where $P_p$ and
$P_{\delta}$ are the corresponding gas pressure power spectrum and
dark matter density power spectrum. This similarity suggests a common
origin. In the hierarchical approach,
$a_4(l)$ is related to $b^2_p(k,z)$ and we 
expect $a_4(l)$ to keep the similar shape but with smaller
amplitude variation than $b^2_p$, as illustrated by the behavior of
$a_4(l)$ in our simulation and $b^2_p(k,z)$ in our analytical model.

We further show the
$y$ PDF (fig. \ref{fig:ydis}) as a function of smoothing
scale. The distribution of
$\log_{10}{y}$ is roughly Gaussian, especially for large smoothing
scales, as suggested by
\citet{Seljak01}. This behavior reflects the non-Gaussianity of the
$y$ parameter, up to angular scales $\sim 10^{'}$. The distribution
of $y$ is asymmetric since the 
distribution of $\log_{10}y$ is nearly symmetric. We quantify it by
the skewness $\Theta_3\equiv \langle 
(y-\bar{y})^3\rangle/\sigma_y^3$ as a function of  smoothing scale. For
symmetric $y$ distribution, $\Theta_3=0$.
Fig. \ref{fig:ngs} shows that $\Theta_3\simeq
10 $ when $\theta\rightarrow 0$. This result again agrees with our
hierarchical method prediction. We predicted, at small angular scales,
$\Theta_3 \sim S_3  \sigma_g^{0.5}(z\sim 1) \sim 10$ where $S_3\equiv
\frac{\langle\delta^3\rangle}{\langle \delta^2  \rangle^2}\sim 5$ 
\citep{Scoccimarro99}. The positive sign of $\Theta_3$ reflects the
numerousness of high $y$ regions, as suggested in fig. \ref{fig:peak}.

 The kinetic SZ effect
is determined by the gas momentum along lines of sight. The
fractional temperature change
$\Theta(\hat{n})=-\sigma_T\int_0^{l_{\rm re}} n_e(l\hat{n}) \frac{{\bf
v}\cdot {\hat{n}}}{c} dl$. Since the
direction of the gas velocity ${\bf v}$ is random, we would expect
$\Theta_3=0$ for a sufficiently large sky. 
The contribution from non-linear structures is partly
cancelled out and thus we expect mildly non-Gaussianity.  As shown in
fig. (\ref{fig:ngs}), the kinetic SZ $\Theta_4$ and $a_4$ are significantly
smaller than the 
corresponding thermal SZ $\Theta_4$ and $a_4$ (fig. \ref{fig:ngs}). The kinetic
SZ effect is 
nearly Gaussian at angular scales $\sim 20^{'}$.

\section{Simulated observations of the SZ effect}
\label{sec:AMIBA}

In real SZ observations, instrumental noise and primary CMB cause
additional errors in the SZ statistics such as the power spectrum and
peak number counts. We need to estimate these effects to derive
optimal observing strategies to measure these statistics in the
presence of noise.  With these observational errors, our methods (\S
\ref{sec:application}) to extract 3D gas information is limited and we
must check their feasibility.  In this section, we take AMIBA as our
target to address these issues.
\subsection{AMIBA}
AMIBA is a 19 element interferometer with $1.2$ meter dishes. All dishes are closely
packed in three concentric rings. It operates at $\nu_{\rm center}=90$
Ghz with $\Delta \nu=16$ Ghz, system noise $T_{\rm sys}=100$ K and system
efficiency $\eta\sim0.7$.  At this frequency, $\Theta\simeq -1.6y$
with $S_T(90 {\rm Ghz})\simeq0.8$.  
The goal of this experiment is to image maps of the CMB sky with arc
minute resolution. We
consider observations with fixed integration time and aim at
finding the optimal sky area $\Omega$ and sky fractional coverage
$f_{\rm sky}=\Omega/4\pi$ for a given statistics. 

For closely packed interferometers observing such weak signals, the
ground fringe can be a major source of noise.  To eliminate the ground
fringe, AMIBA plans to drift scan: the telescope is parked while the
sky drifts by.  This assures that the ground fringe remains constant
with time.  The mean value of each fringe is then subtracted from the
scan, cleanly eliminating the ground.  A field is mosaicked by a
series of adjacent scans, and the most uniform coverage is achieved by
incrementally offsetting the pointing center on each scan to yield a
finely sampled 2-D map.  The raw output of the experiment are
correlations, two for each baseline, polarization and frequency
channel, corresponding to the real and imaginary correlator outputs.
We can think of each of these outputs to correspond to an image of the
sky filtered through some anisotropic beam.  As a first step, we can
combine degenerate baselines and polarizations, reducing the 171
baselines to 30 non-degenerate baselines.  Since the CMB is expected
to not be significantly unpolarized, we can combine the two
polarization channels, leading to 
60 raw maps per frequency channel.  These maps can be merged optimally
into one global map by convolving each map with its own beam, and
scaling each map to the same noise level, and coadding these maps,
resulting in a 'clean' map.  Each of the constituent maps had
explicitly white noise, so the noise statistics of the summed map are
easily computable.  The 'clean' map can be deconvolved by the natural
beam, resulting in a 'natural' map.  This 'natural' map is an image of
the sky convolved with the natural beam of the telescope plus white
noise.  The angle averaged natural beam is shown in fig. \ref{fig:beam}.  The CMB intensity fluctuation $\delta I_{\nu}$ measured
by AMIBA has three
components: the primary CMB, the SZ effect and the 
instrumental noise. It is related to the temperature fluctuation by
\begin{equation}
\frac{\delta I_{\nu}}{I_{\nu}^{\rm CMB}}
=\frac{x \exp(x)}{\exp(x)-1}\left[\left(\frac{\delta T}{T}\right)_{\rm
CMB}+\left(\frac{\delta
T}{T}\right)_{\rm SZ}+\frac{\delta T^N}{T_{\rm RJ}}\right]
\end{equation}
Here, $\left(\frac{\delta T}{T}\right)_{\rm CMB}$ and $\left(\frac{\delta
T}{T}\right)_{\rm SZ}$ are the corresponding temperature fluctuations seen through
the AMIBA natural beam. $x=h\nu/k_B T_{\rm CMB}$. We have chosen the
normalization of the beam 
such that the noise power spectrum is white with  
\begin{equation}
C_N=4 \pi
T^2_{\rm sys}f_{\rm sky}/(2\Delta \nu t \eta^2)/T_{\rm RJ}^2.  
\label{eqn:noise}
\end{equation}
$T_{RJ}=T_{\rm CMB} x^2 \exp(x)/[\exp(x)-1]=2.22$ K is
the Raleigh-Jeans equivalent CMB temperature at $\nu=90$ Ghz.  The
factor of $2$ is due to the two polarizations of the AMIBA experiment.
For a single dish of infinite aperture, with a single pixel detector,
the window would be identically one: the scanned image is just the CMB
distribution on the sky.  Since AMIBA has many detectors, one can
combine them to either lower the noise, or to boost the signal.  We
have chosen to use the equivalent noise of a single pixel single dish
experiment, and normalized the beam accordingly. We calculate the
natural beam $W_N(l)$ by equation (17) and (46) of \citet{Pen02}. The
natural beam in multipole space and real space are shown 
in fig. \ref{fig:beam}.  The natural beam has a FWHM of $2^{'}$. It 
peaks at $l_i\simeq 2\pi \lambda/D_i$. Here, $\lambda\simeq 3.3$ mm is
the AMIBA operating wavelength and $D_i$ is the distance of
the $i$th baselines. The first peak $l^{\rm peak}_1\sim 2273$ corresponds
to the shortest base line $D_1=1.2$ m. At this angular scale, the sum
of all the baselines improves throughput by a factor of almost three
($\epsilon\simeq 2.77$).  This is
analogous to having a nine pixel detector on a single dish.  More
detailed definitions of the beams and strategies are described in
detail in \citep{Pen02}.

\begin{figure}
\plotone{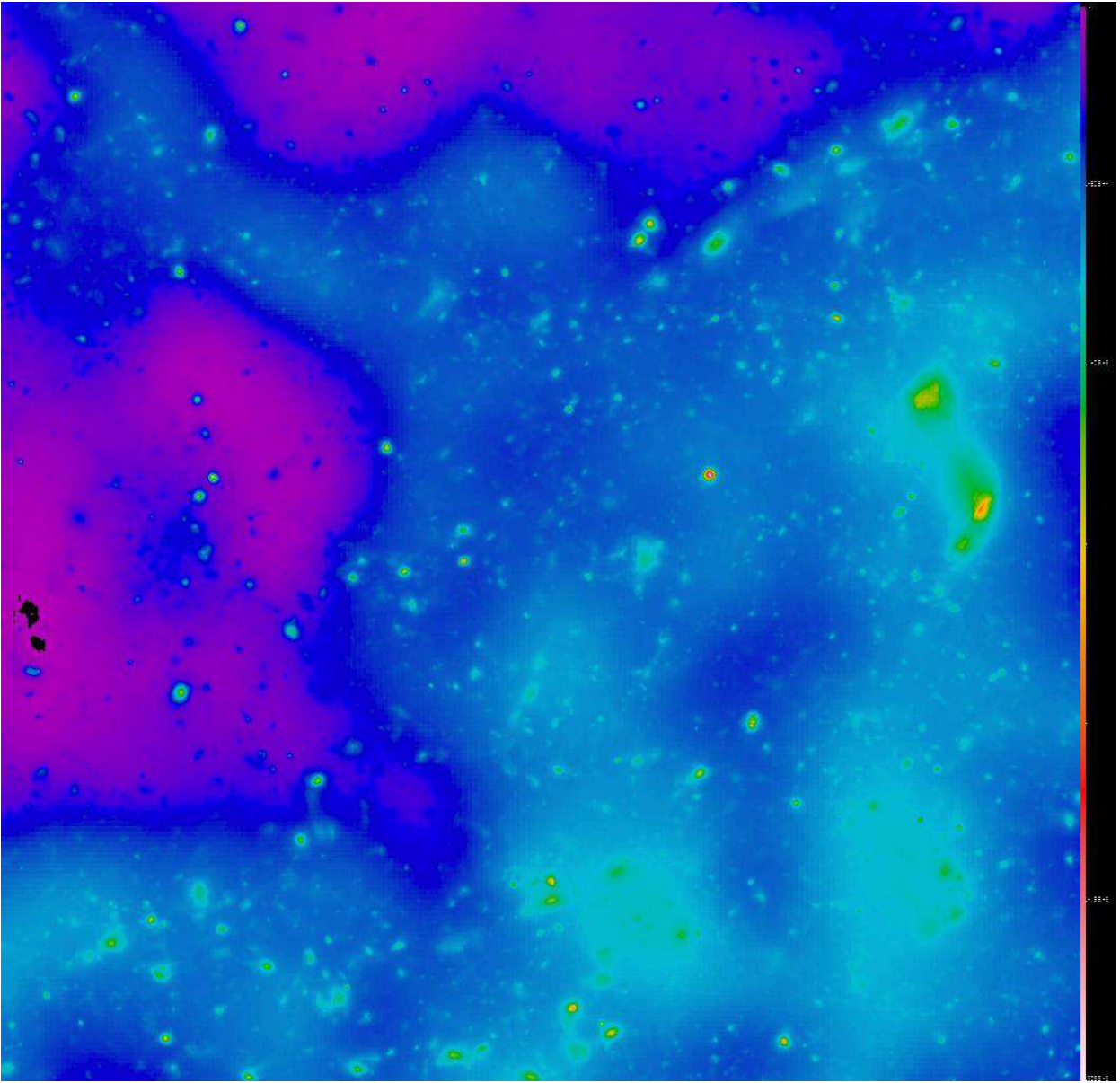}
\caption{A combined SZ+CMB map. At large angular scales, the primary
CMB dominates and smears the structure of the thermal SZ
effect. \label{fig:tc}} 
\end{figure}

In our simulated pipeline, we add primary CMB map fluctuations to
our simulated sky maps using CMBFAST-generated \citep{CMBFAST} power
spectra with same cosmological parameters except for the use of COBE
normalization for $\sigma_8$, which is  slightly different from our
simulation value.
As we will show below, the primary CMB is not  
the main source of noise for the SZ power spectrum measurement at
AMIBA angular scales ($l>2000$) and  is negligible in SZ
cluster searches. So the effect of this inconsistent $\sigma_8$ is
insignificant in our
analysis. Adding SZ maps and CMB maps, we obtain simulated sky
maps (fig. \ref{fig:tc}). In these maps, SZ  structures, especially those  
caused by diffuse IGM are superimposed with the primary CMB. We then
convolve these maps with the  natural beam. The beam function decreases
quickly to zero towards large angular scales where the primary CMB
dominates, so it efficiently filters most primary CMB structures
larger than this scale.  We then add the noise given
by eqn. (\ref{eqn:noise}) to this map.   The normalization
in the beaming and  filtering (described below)  process is
arbitrary. We  choose the normalization in such a  way that the power
spectrum of the map at the scale of the peak response does not change
after beaming or 
filtering.  It corresponds to normalizing the global maximum of the
beam function to be unity. Under such normalization, when we add
instrumental noise to the simulated CMB+SZ map, the noise is depressed
by a normalization factor $\epsilon\simeq 2.77$. Our $1.19^{\circ}$ map has
$2048^2$ pixels, so the white instrumental noise dominates on small
scales.  For a $20$ hours$/{\rm deg}^2$ survey, the 
dispersion of the noise temperature fluctuation  
$\sigma_N=T_{\rm sys}/\sqrt{(2\Delta \nu t_{\rm pixel}}/(\eta
T_{\rm RJ} \epsilon) \sim 0.002\gg \bar{y}$. $t_{\rm pixel}$ is the observing
time for each pixel. So all signals are hidden under the instrumental
noise. One needs further filtering to obtain an image
that is not overwhelmed by the small scale noise.  Ignoring the CMB
fluctuations, we know that point sources have the shape of the beam.
A point sources optimized search would convolve the natural map with
the shape of the beam, and peaks in this map correspond to the maximum
likelihood locations of point sources.  We can think of the CMB as a
further source of noise, and filter that away as well.  In this case,
the filter depends on the ratio of CMB and noise amplitudes, i.e. the
actual integration time.  One can also optimize a filter for a
structure of a known intrinsic shape.  Since clusters are
approximately isothermal, a better filter would be one matched to the
extended isothermal nature, where objects in the natural map have the
shape of an isothermal sphere convolved with the natural
beam. Combining all these consideration, 
the optimal filter for noise cleaning is given by $W_C(l)=W_F(l)
W_N(l)/(1+W^2_N 
C_l^{\rm CMB}/C_N)$ (fig. \ref{fig:beam}).  $W_F(l)$ is the Fourier
transform of the source intrinsic shape $W_F({\bf r})$. For point sources,
$W_F(l)=1$. For clusters, on scales smaller than the cluster virial
radius and larger than the core radius, 
$y(\theta)\propto \theta^{-1}$ ($\theta$ is the angular distance to
the cluster center). Then
$W_F(l)\propto l^{-1}$.  Since the corresponding angular size of the
core radius is much smaller than the beam size, the above
approximation is sufficiently accurate. We shown the filter with
$280$ hours per square degree scan rate in figure
\ref{fig:beam}. It has a FWHM of $2^{'}$.  In
multipole space, it  drops  to near
zero at angular scales $l \sim 1500$ and $l>9000$ and peak at cluster
scales $l\sim 3000$.  So it is efficient to filter away the
residual primary CMB and the 
instrumental noise while amplifying SZ signals. 
We tried different sky scan rates to find the optimal survey
strategy.  We show final
resulting maps in fig. (\ref{fig:filter}) with $280$
hours per square degree scan rate.  
With the simulations we have the luxury of seeing maps with (right
panels of fig. \ref{fig:filter}) and
without noise (left panels of fig. \ref{fig:filter}). We discuss the
simulated AMIBA measurement of the 
SZ power spectrum and cluster searching in the next two subsections.

\begin{figure}
\plottwo{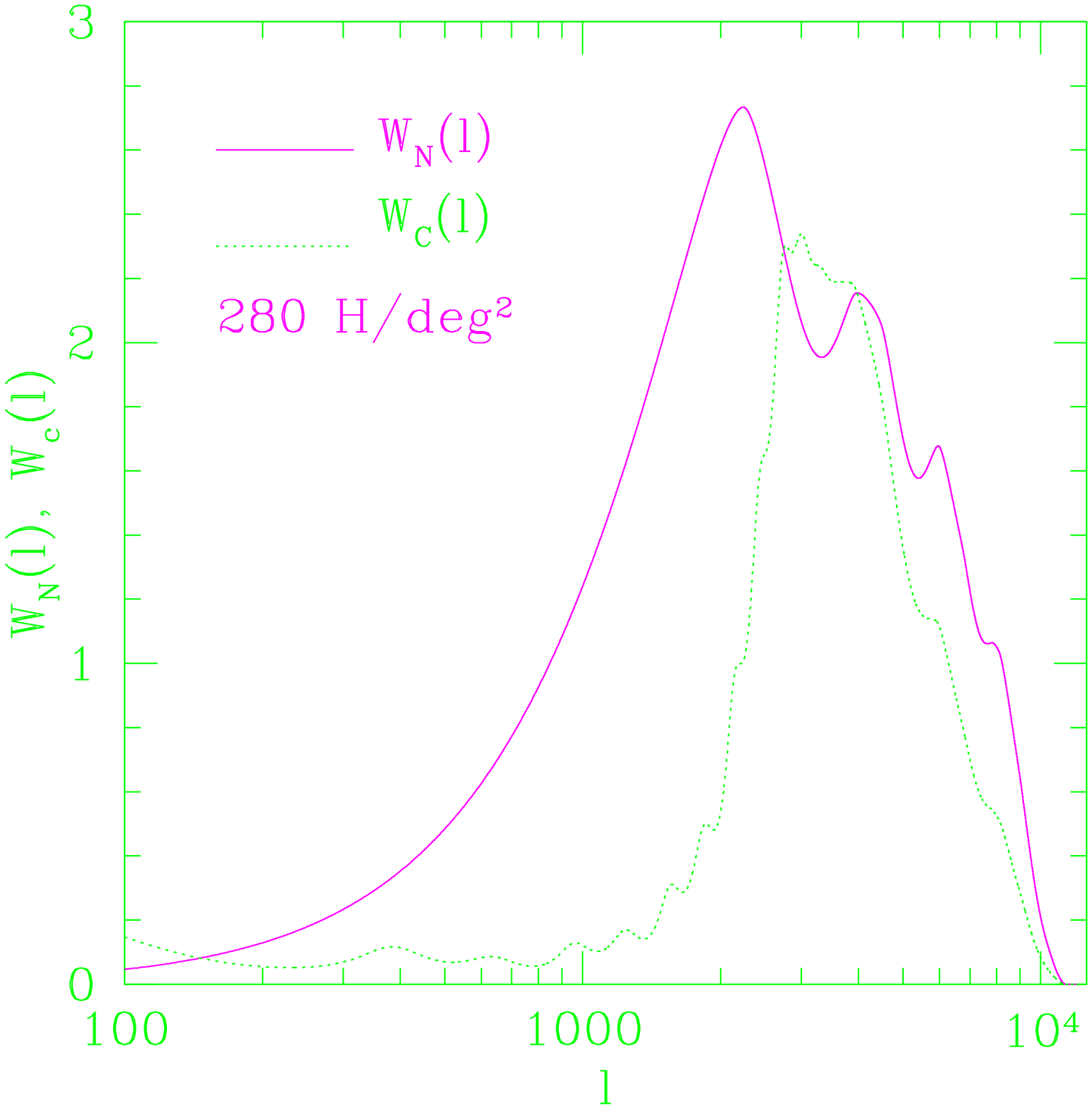}{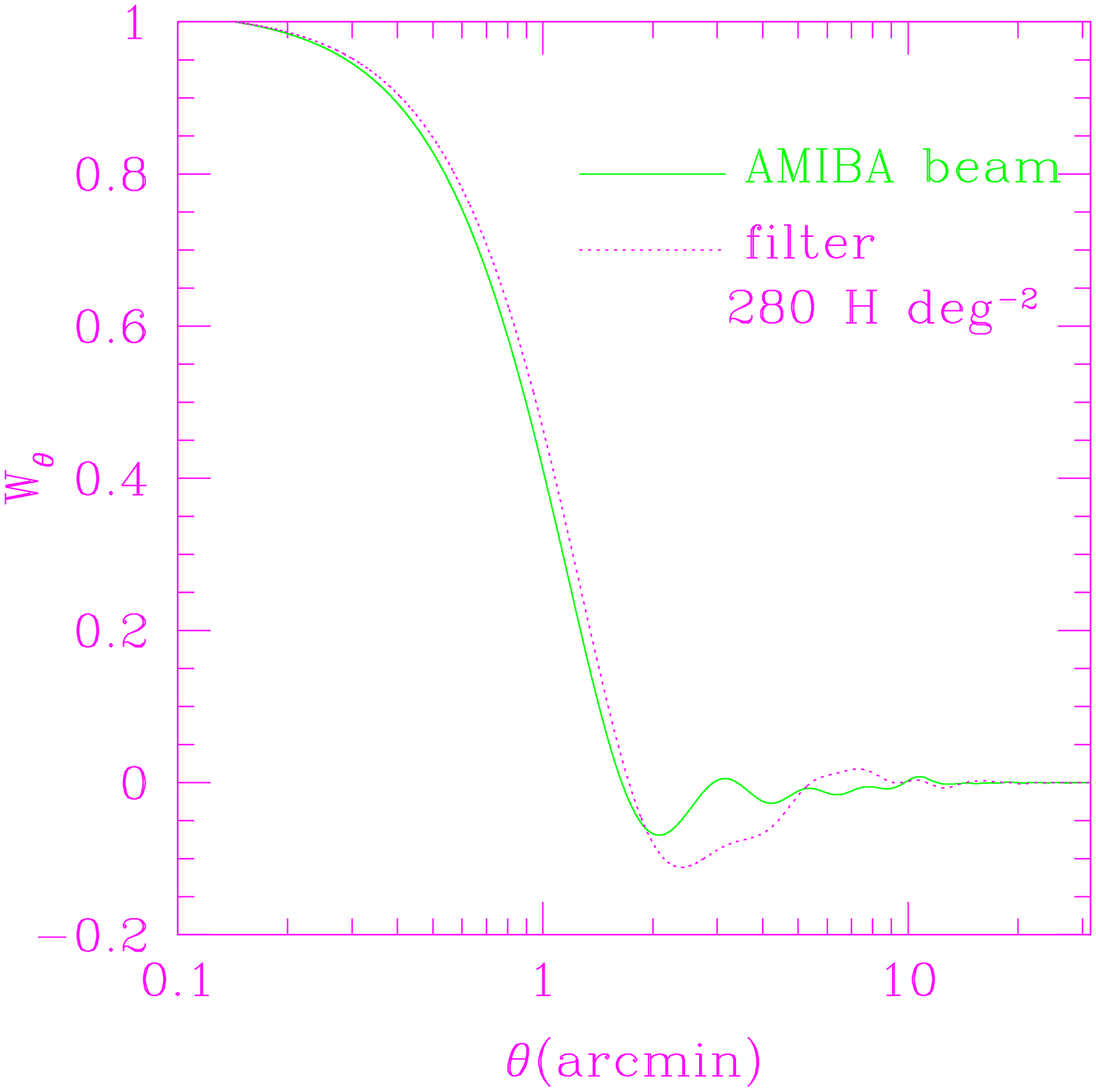}
\caption{The natural beam and optimal filter functions of AMIBA in multipole
(left panel) and real space (right panel). The main
goal of the beaming and filtering is to filter away the primary
CMB, which dominates at large scale ($l\lesssim 1000$) and
instrumental noise at small scale($l\rightarrow \infty$). This goal is clearly
illustrated in the large and small $l$ behavior of these
functions. The optimal filter depends on the noise amplitude. We show
the case for AMIBA scan rate $280\ {\rm hours}\ {\rm deg}^{-2}$. This
filter has a $\sim 2^{'}$ FWHM. \label{fig:beam}}  
\end{figure}
\begin{figure}
\plottwo{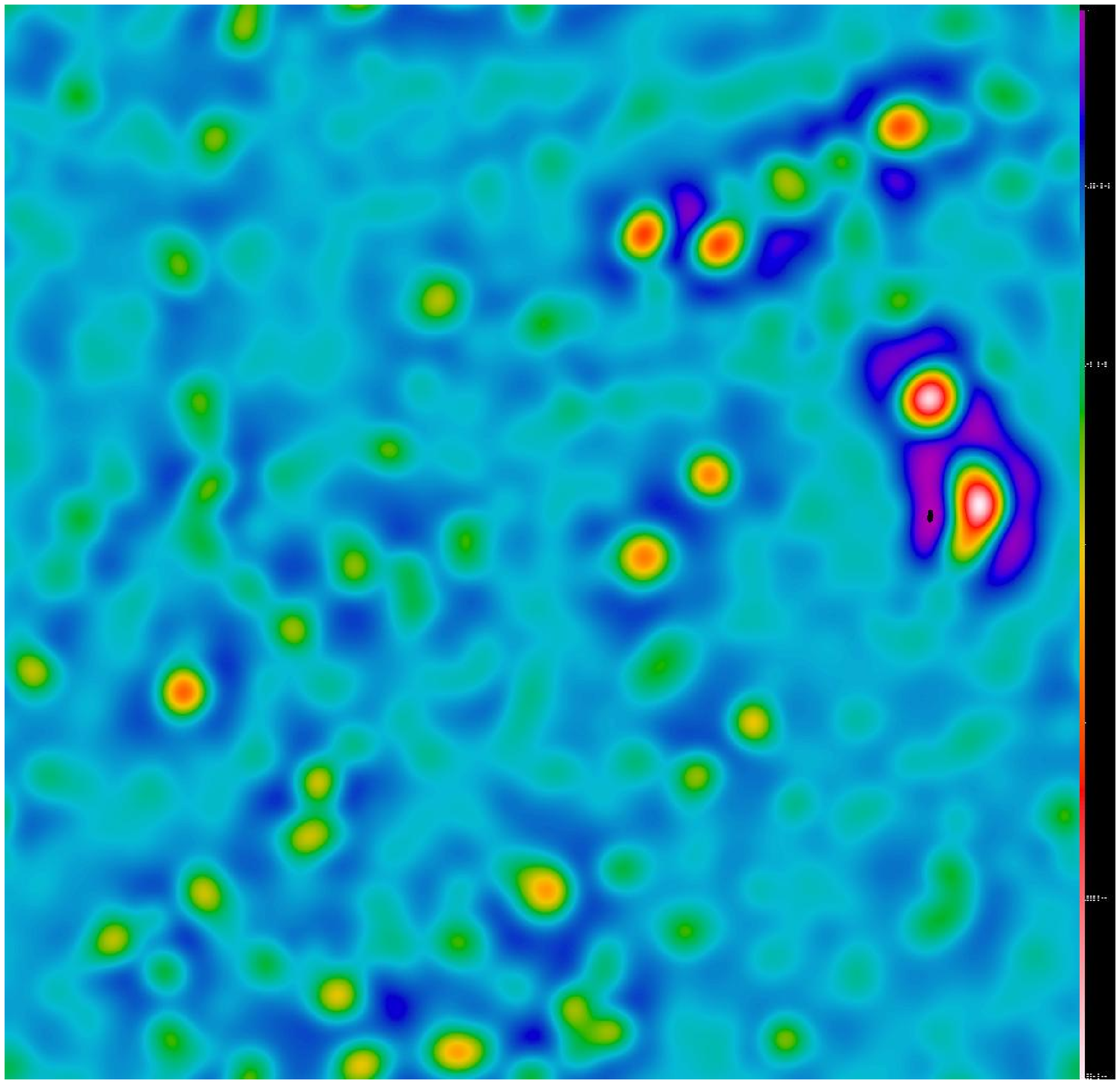}{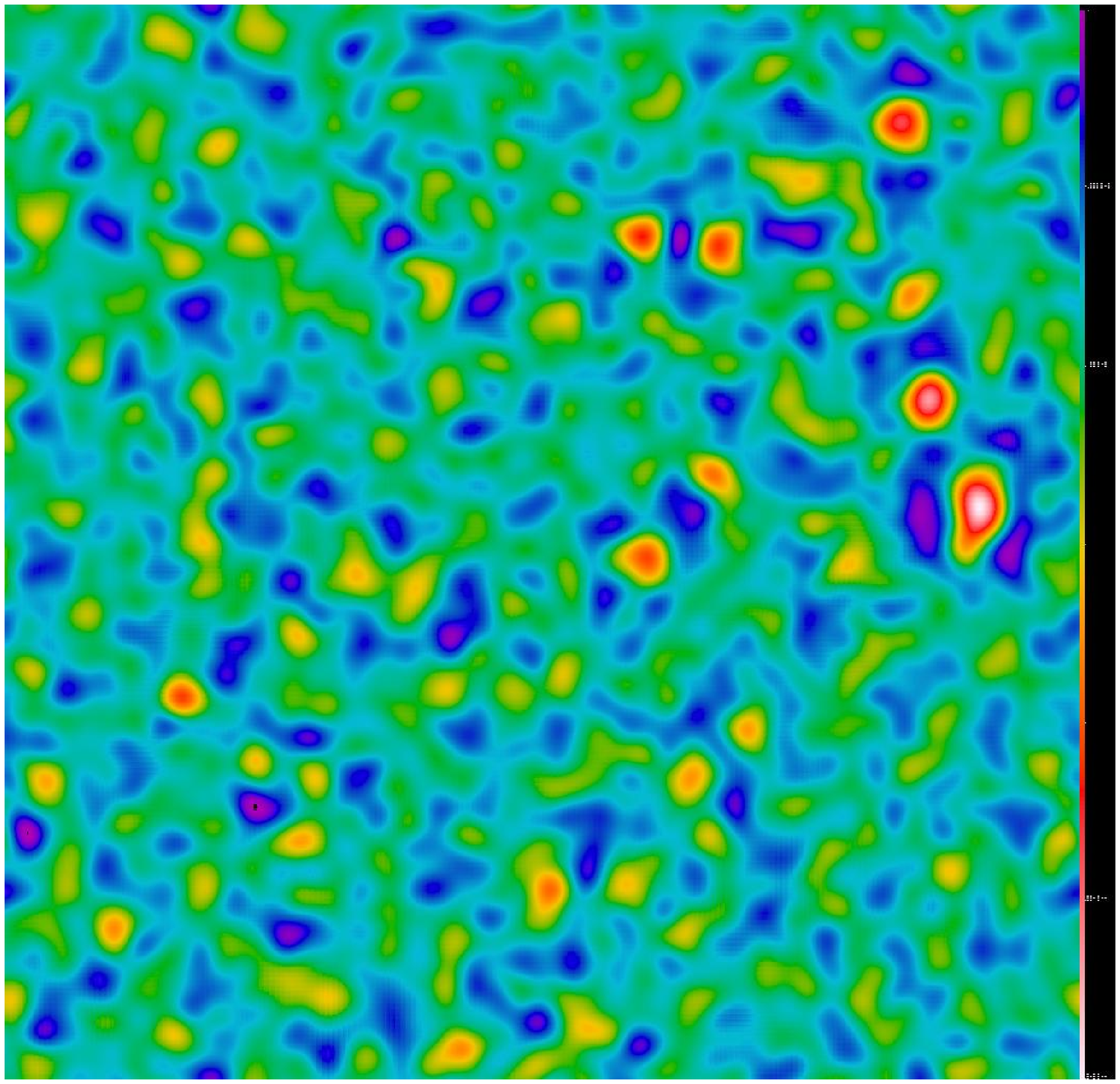}
\caption{Filtered maps without noise (left panel) and with noise
(right panel) of AMIBA $390$ hours observing. Many peaks in the total map
are not real signal peaks while some
peaks in the clean map disappear in the total map.  For AMIBA's
frequency, $\Theta=-1.6 y$. For clarity, we plot $-\Theta$.
\label{fig:filter}} 
\end{figure}
\begin{figure}
\plottwo{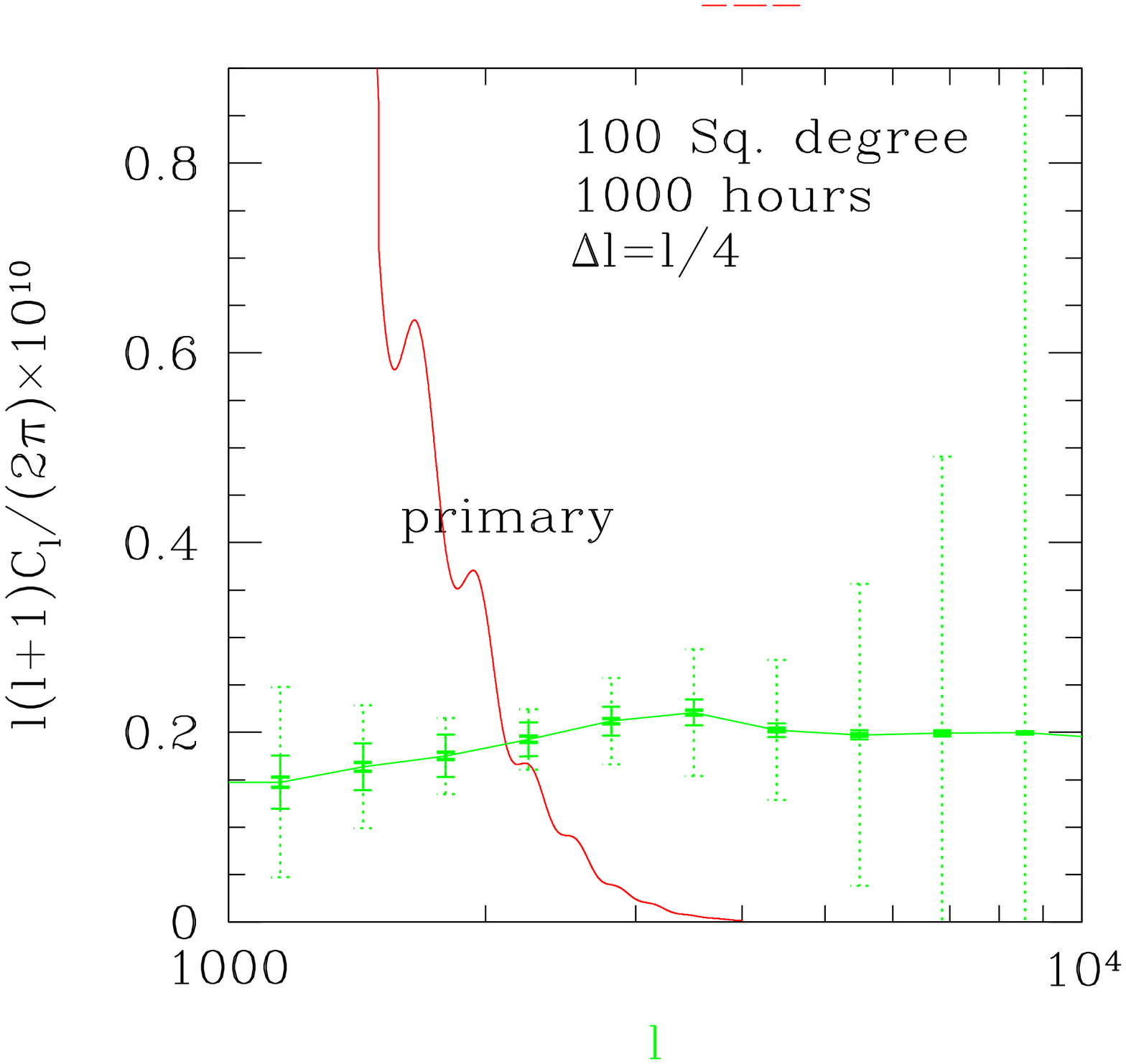}{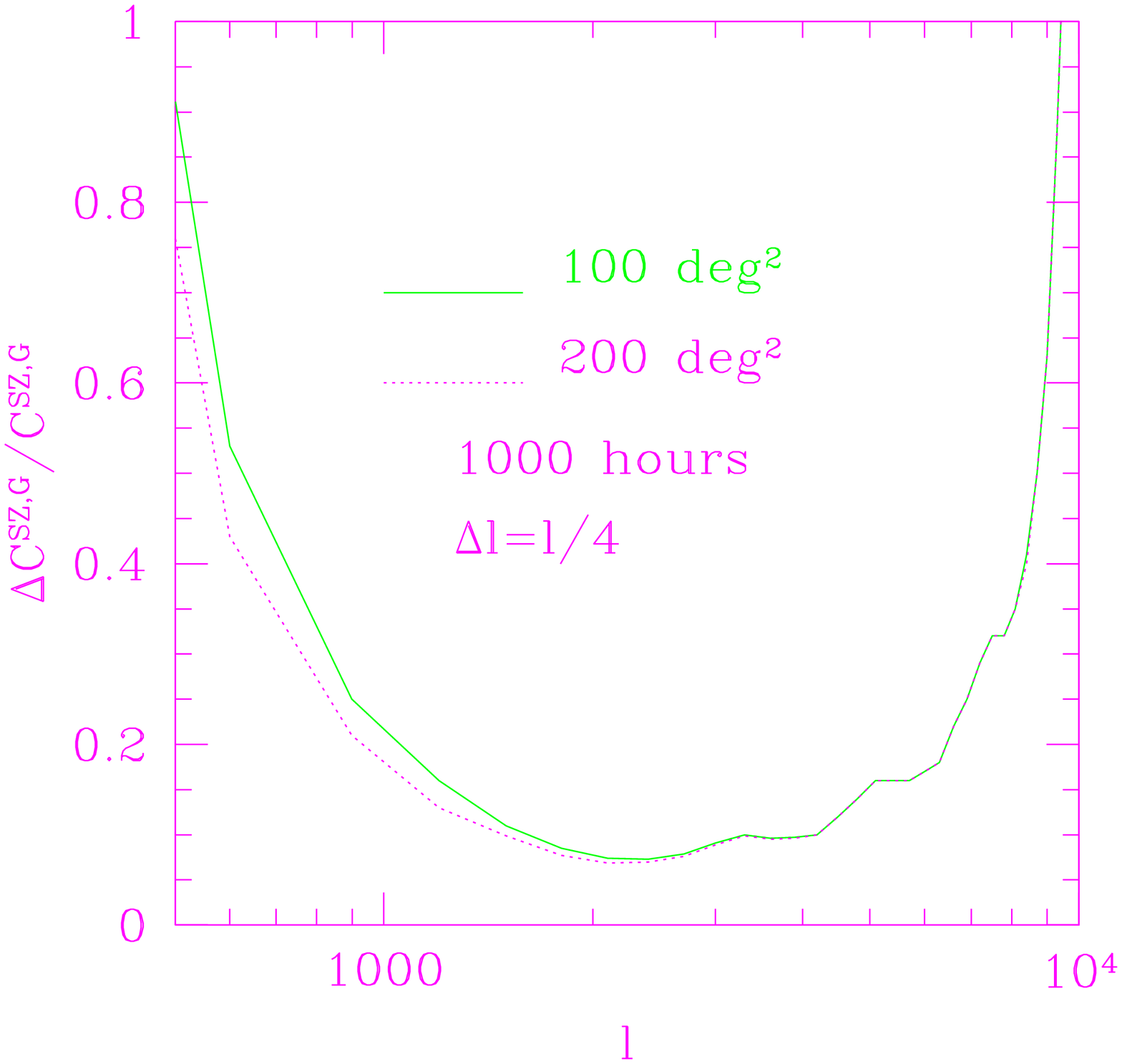}
\caption{Errors in the AMIBA measurement of SZ power spectrum (left
panel) and
cross correlation (right panel) with SDSS galaxies.  The SZ power spectrum is
$0.8^2$ times of the one in fig. \ref{fig:cl} since at
AMIBA operating frequency $\nu=90$ Ghz, $\Theta=-1.6 y$ as contrast to
$\Theta=-2 y$ in the  Rayleigh-Jeans regime . The thick solid
error bar is  the Gaussian variance, the thin solid error bar is the
actual variance calculated from a $40$ map ensemble and
the dot error bar is the total error including  primary CMB and
instrumental noise.  For a $1000$ hours survey of a $100\ {\rm deg}^2$
area, the accuracy in the SZ power spectrum measurement is about $40\%$
at the range of $2000< l< 5000$. In the same survey, the measured
cross correlation has about $20\%$ accuracy at
similar $l$ range. \label{fig:error}} 
\end{figure}
\begin{figure}
\plotone{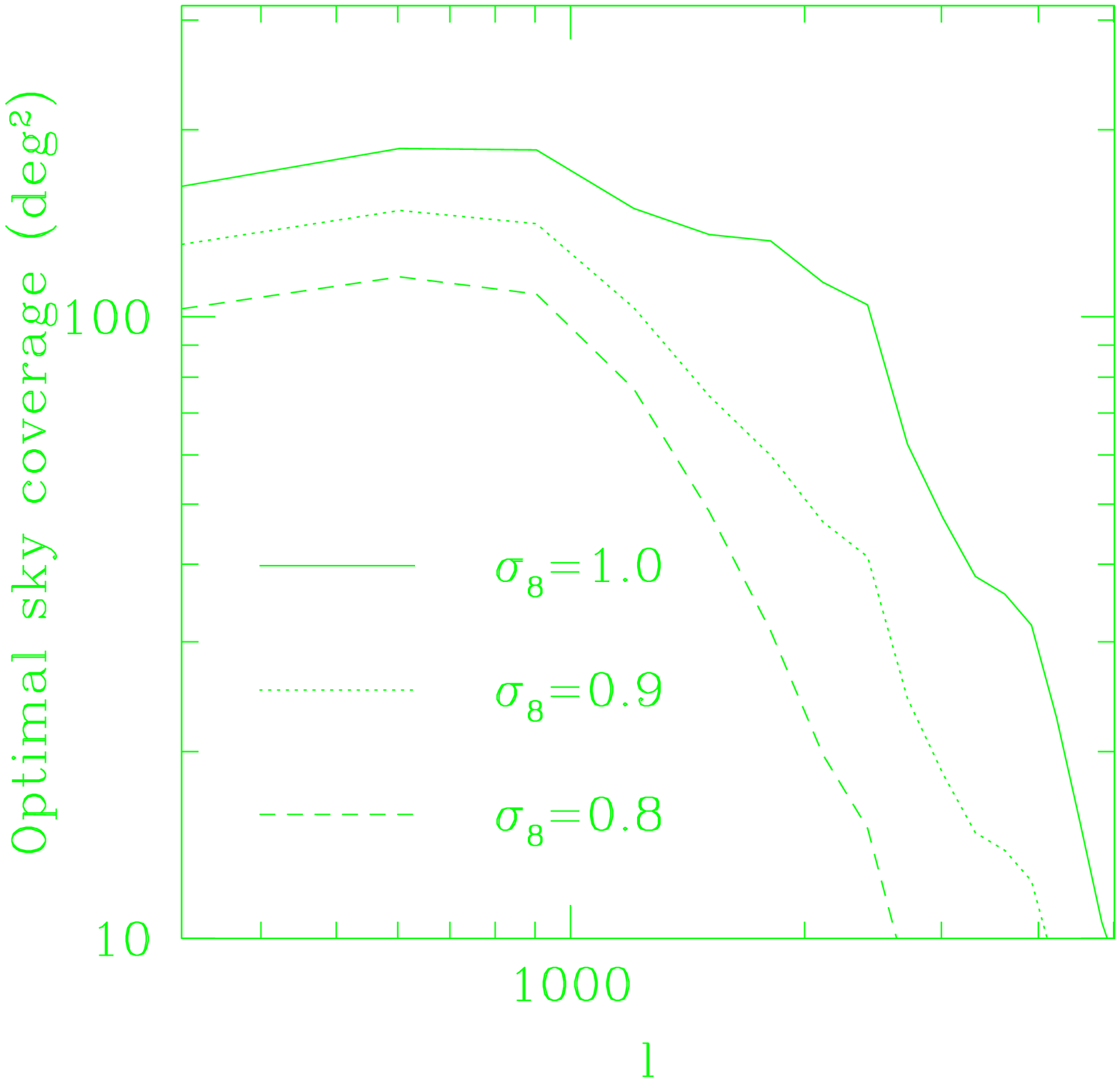}
\caption{The dependence of the optimal sky coverage in the SZ power
spectrum measurement on $\sigma_8$ in a $1000$ hour AMIBA survey. Our
simulation used $\sigma_8=1.0$. The $\sigma_8=0.9$ and $\sigma_8=0.8$
cases are estimated assuming the strongest dependence of $C_l$ on $\sigma_8$:
$C_l \propto \sigma_8^9$. \label{fig:fsky}}
\end{figure}

\subsection{The SZ power spectrum in the simulated AMIBA experiment}
\label{subsec:power}
One of the key goals of AMIBA is to measure the SZ power spectrum. As
discussed in \S \ref{sec:simulation}, it is a sensitive measure of the
gas thermal history. Furthermore,  combining the  cross-correlation with
photometric galaxy surveys, an SZ survey can measure the IGM pressure
power spectrum as a
function of redshift.  This gives us access to the evolution of the
IGM state. For the purpose of the 
SZ power spectrum estimation, the two noise sources, the primary CMB
and the thermal instrument noise, are both Gaussian. The intrinsic SZ variance
causes further error. The combined error for the power spectrum
estimation is 
\begin{equation}
\Delta C^{\rm SZ}_l=\sqrt{\frac{a_4(l)
C^2_{\rm SZ}(l)+2\left[C_{\rm SZ}(l)+C_l^{\rm CMB}+C^N_l/W_N^2\right]^2}{(2 
l+1)\Delta l  f_{\rm sky}}}. 
\end{equation}
 If the SZ effect is Gaussian
($a_4=0$), we recover the usual expression of the error. We take a large bin 
width $\Delta l=l/4$ to estimate the error. Since the SZ power spectrum is
nearly flat in the range  $l \sim 
2000-15000$, this choice of bin size does not lose significant
information.
At large angular scales ($l<1500$), the error from primary CMB
dominates, and 
at small scales ($l>5000$), the instrumental noise dominates. In the
intermediate scales, the intrinsic error of the SZ effect dominates.
Different errors have different dependences on the sky
coverage. Increasing $f_{\rm fsky}$ while fixing the integration time, the errors from primary CMB and SZ
effect decrease but the the one from the instrumental noise increases,
so there exists a optimal sky coverage for a given  integration time. 
Since the errors from primary CMB and SZ effect are both scale as $
f_{\rm sky}^{-1/2}$ and the instrumental noise scales as $f_{\rm
sky}^{1/2}$, the minimum error is obtained when $(a_4+2) 
C^2_{\rm SZ}+2C_l^{\rm CMB}+4C_{\rm SZ}C^{\rm CMB}=2C_N^2/W_N^4$, which gives the
optimal sky coverage. For $1000$ hours of  
observing, several hundred square degree sky coverage  is nearly
optimal (fig. \ref{fig:fsky}).   Fig. \ref{fig:error} shows that
AMIBA is able to 
measure the SZ power spectrum with an accuracy of $\sim 40\%$ for $l$ between
$2000$ and $5000$ in $1000$ hour observing of $100$ square degrees of
sky. The strong dependence of the SZ effect
on $\sigma_8$ strongly affects 
our error estimation. A smaller $\sigma_8$ reduces the signal and
S/N. In order to keep S/N, smaller
sky coverage is needed to reduce the noise. The dependence of the optimal sky
coverage $f_{\rm sky}^{\rm opt}$ on $\sigma_8$ is shown in
fig. \ref{fig:fsky}.  In this estimation, we have assumed the
most extreme dependence of the SZ effect on $\sigma_8$, namely,
$C_{SZ}(l)\propto \sigma_8^9$.

If we cross correlate the observed SZ effect with SDSS, we can
extract the underlying 3D 
gas pressure power spectrum and pressure-galaxy cross correlation. To
test its feasibility, we estimate the error in the angular cross
correlation measurement by
\begin{equation}
\frac{\Delta C^{\rm SZ,G}(l)}{
C^{\rm SZ,G}(l)}=\sqrt{\frac{1+r^{-2}(1+C^{\rm CMB}/C_{\rm
SZ}+C^N_{\rm SZ}/C_{\rm SZ})(1+C^{N}_G/C^G)}{(2l+1)\Delta
l f_{\rm sky}}}.
\end{equation}
SDSS will cover $f^{\rm SDSS}_{\rm sky}=1/4$ of the sky and
will detect $N_G\simeq 5\times 10^7$ galaxies with photometry in five
bands. The Poisson noise in SDSS has the power spectrum $C^N_G=4\pi
f^{\rm SDSS}_{\rm sky}/N_G$. We assume a linear bias between galaxy number
overdensity and dark matter overdensity to calculate the galaxy surface
density power spectrum $C^G(l)$. Since SDSS is flux limited,
we take the galaxy selection function
$\frac{dn}{dz}=3z^2/2/(z_m/1.412)^3\exp[-(1.412z/z_m)^{3/2}]$
\citep{Baugh93} with a SDSS fit $z_m=0.33$ \citep{Dodelson01}. The galaxy-SZ
power spectrum $C^{SZ,G}(l)$ is estimated by the SZ-galaxy cross
correlation coefficient $r\equiv C^{\rm SZ,G}/\sqrt{C^{\rm SZ}C^G}$. We choose
$r=0.7$ \citep{Zhang01a} as 
predicted by the hierarchical model. We show the error in
fig. \ref{fig:error}. The larger the sky coverage, the smaller the
error. With the optimal scan rate for SZ power spectrum measurement,
the accuracy in the cross correlation measurement is about
$20\%$. 

%Since $z_m=0.33$ is the median redshift of SDSS galaxies in
%the band $20<r^*<21$ and SDSS galaxies are mostly brighter and closer galaxies
%with $18<r^*<20$, the actual SDSS galaxy auto correlation function
%may have a larger amplitude than our estimation, which increases the
%signal to noise ratio. So, our result of $\frac{\Delta C^{\rm SZ,G}(l)}{
%C^{\rm SZ,G}(l)}$ would be a lower limit of the cross correlation
%measurement accuracy. 

\subsection{AMIBA cluster search}
\begin{figure}
\plottwo{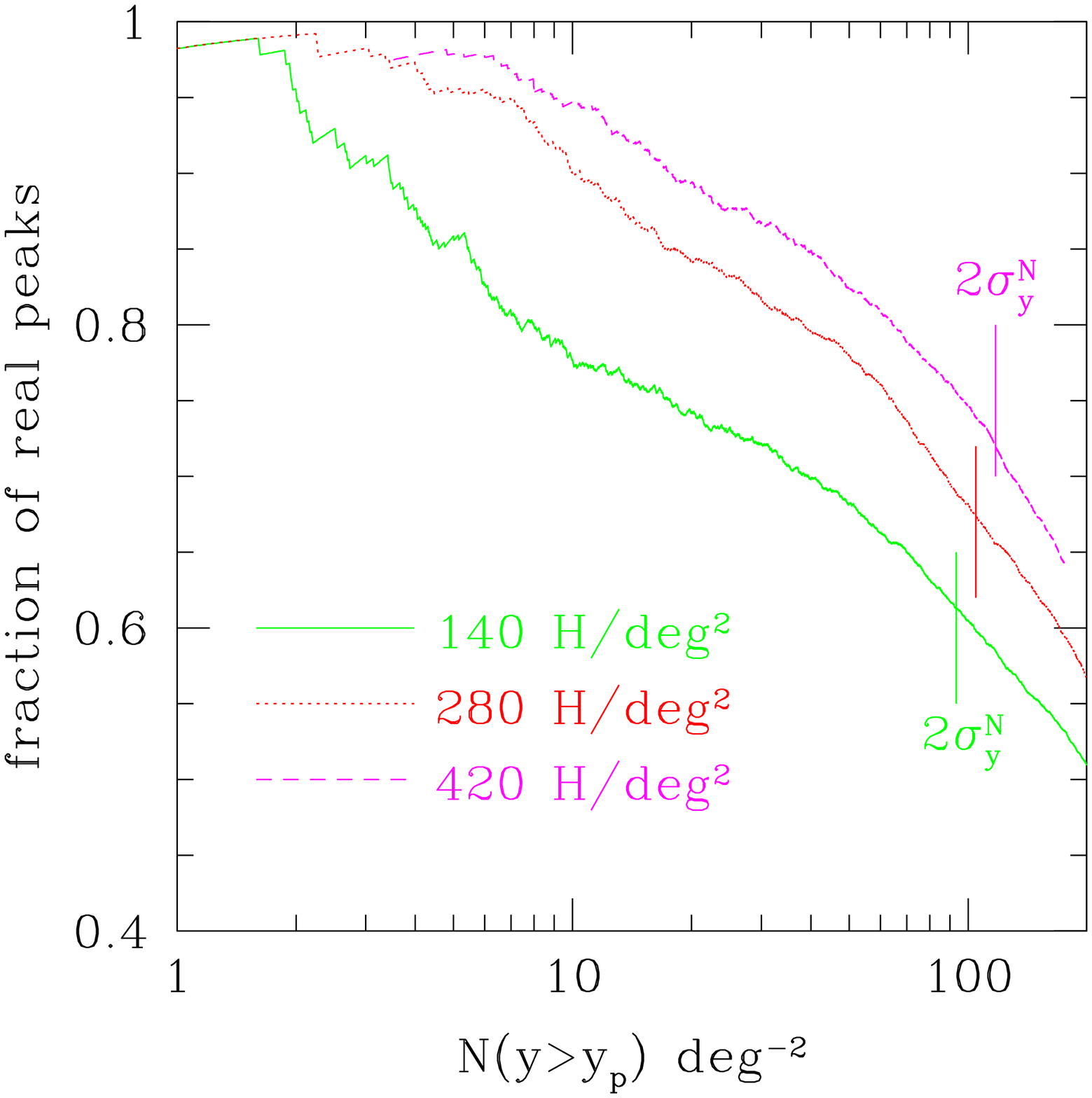}{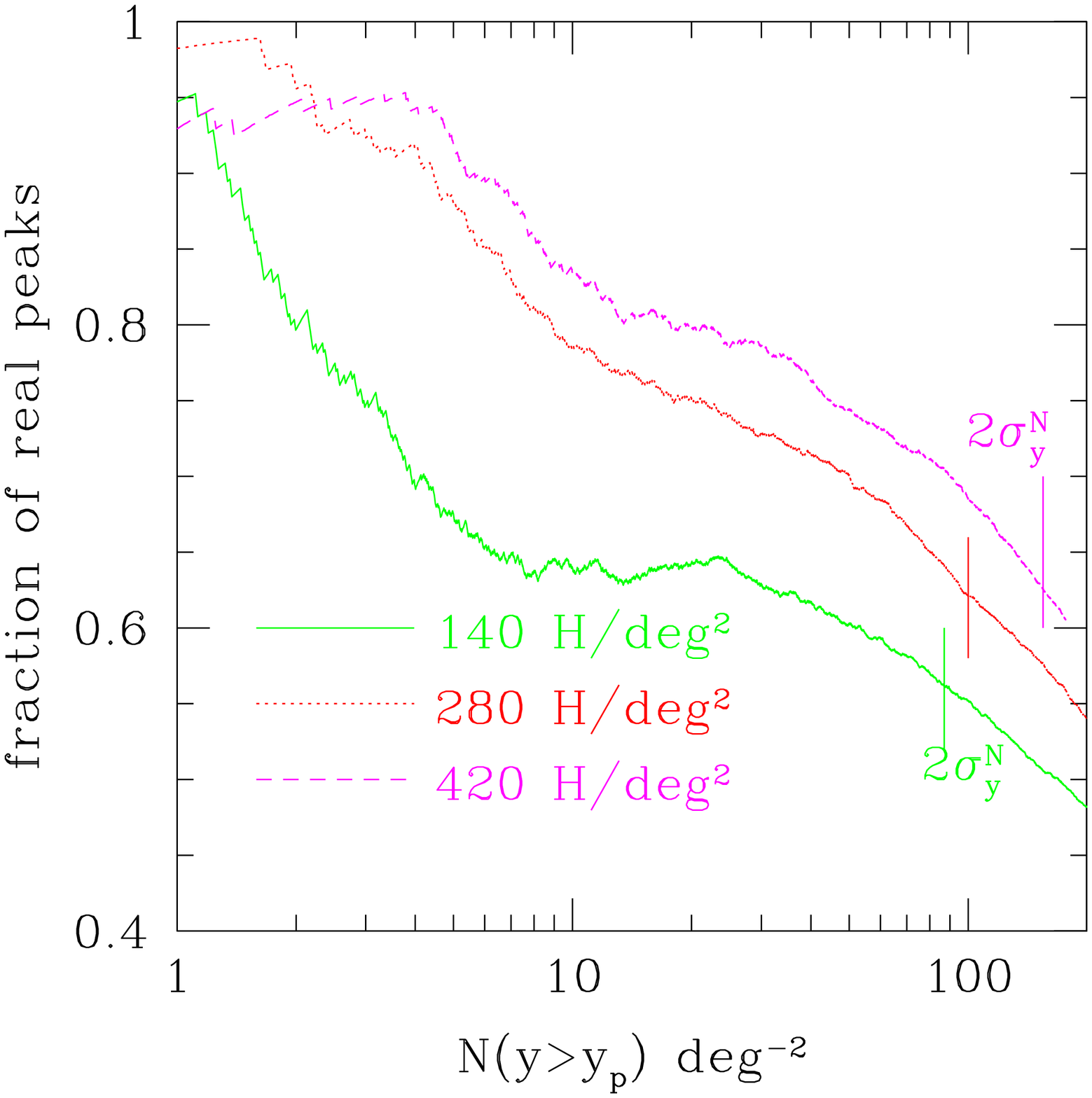}
\caption{The fraction of real peaks as a function of  number of peaks
in simulated AMIBA observing (left panel) with $\sigma_8=1.0$ and the
expected result  
scaled for a different $\sigma_8=0.9$ (right panel). In our
simulation, the optimal cluster 
searching rate is $\sim 1$ cluster every 7 hours, which drops to $\sim 1$
cluster every 30 hours for $\sigma_8=0.9$. We have assumed the extremest $y$
dependence on $\sigma_8$: $y\propto \sigma_8^3$. \label{fig:frac}} 
\end{figure}

\begin{figure}
\plotone{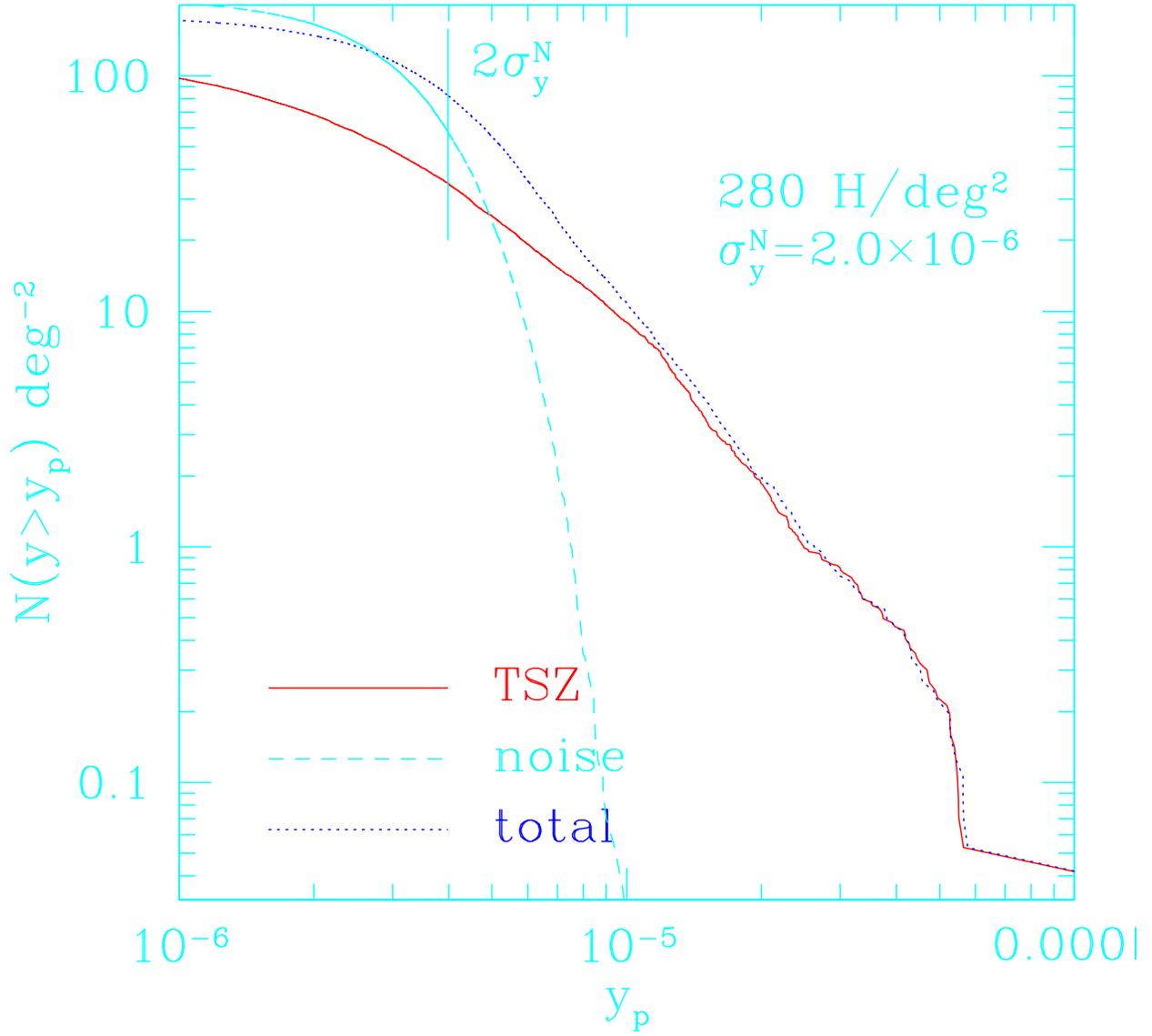}
\caption{Peak distribution $N(y>y_p)$  in simulated
AMIBA observing. All results are averaged from $40$ maps. the 'TSZ' line  
is the pure SZ peak distribution without noise and the 'total' line is what
would be measured in our simulated AMIBA experiment. We see that when
$y<2\sigma_y^N$, peaks due to noise dominate the peak CDF and makes
the measurement of peak CDF unreliable at this region. 
\label{fig:finalpeak}}
\end{figure}

\begin{figure}
\plotone{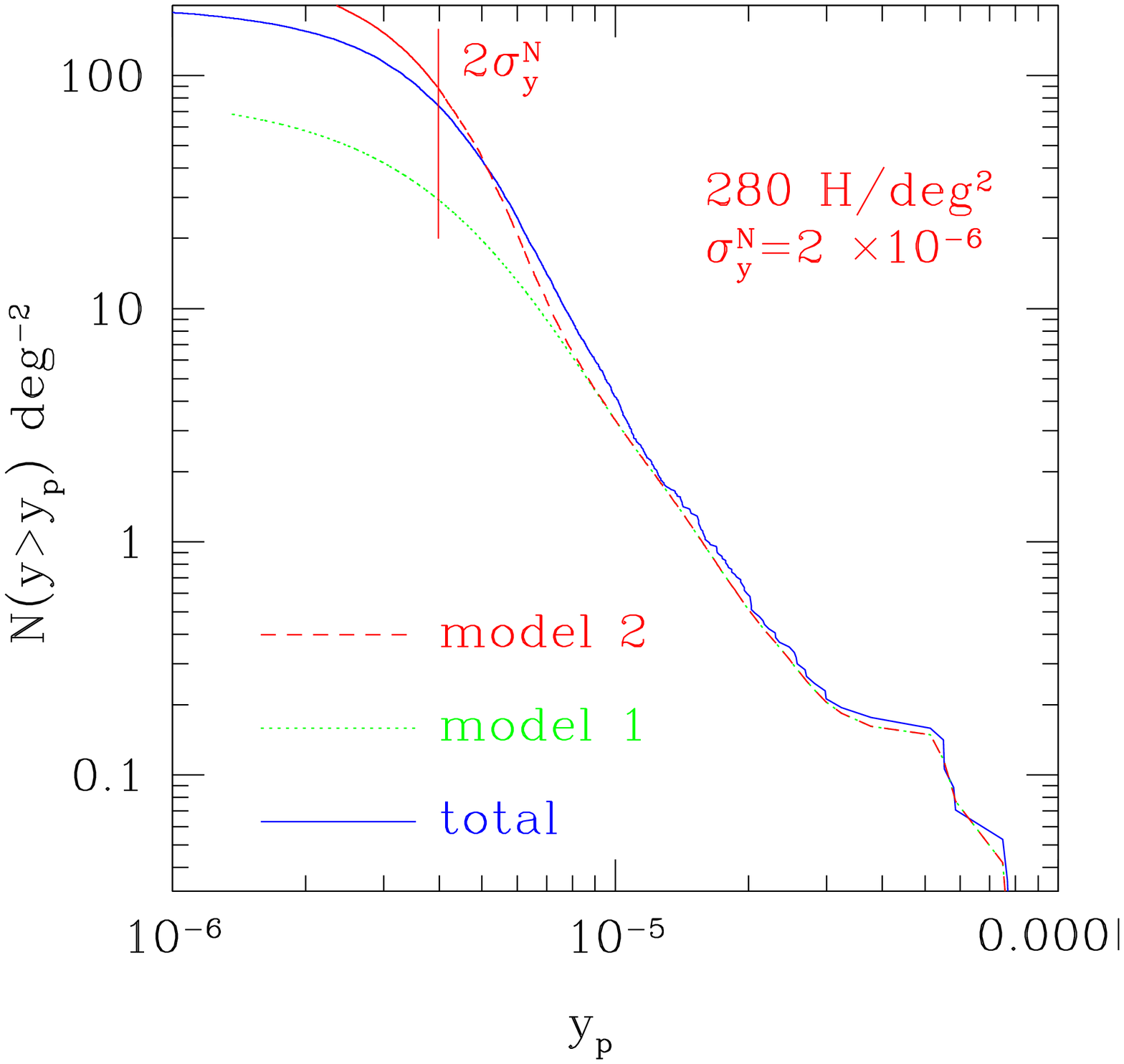}
\caption{Modelled $N(y>y_p)$. The line 'model 1' is calculated
from eqn. (\ref{eqn:ntot}), which consider the effect that noise changes
the amplitude of peaks. The line 'model 2' is the sum of
'model 1' and noise peak CDF, which considers the effect that
noise introduces false peaks. This result fits  the CDF of
peaks in the total maps down to $y_p\simeq 2 \sigma_y^N$ with better
than $30\%$ accuracy. \label{fig:recover}}
\end{figure}

Another important goal of AMIBA is searching for clusters.  In the power
spectrum measurement, the observable is the direct sum of signal and
noise, so the noise contribution can be subtracted linearly in the
power spectrum.  The only net effect is an increase in the error
bars, which is important only at very small angular scales
or very large scales.
But when counting peaks in a map, effects of noise are much more
complicated and are not easily interpreted.  Noise introduces false
peaks in the observed SZ maps, changes the value of real peaks, 
shifts the peak positions and even makes real peaks disappear.  So noise
affects the measurement of cluster 
counts, richness and positions.  Thus we would want a much longer
integration time for the purpose of cluster search. We count peaks
in the filtered maps with and without noise and quantify
these effects as follows. (1) We estimate the accuracy of $y$
measurements by calculating 
the $y$ dispersion of noise $\sigma^N_y$. 
(2) We distinguish real peaks from false peaks by 
comparing the position and amplitude of each  peak in the clean SZ maps
without noises and in the 
total maps with noises (fig. \ref{fig:frac}). One FWHM is
roughly the size of 
noise structures after filtering and corresponds to the maximum position shift
noise can exert to real peaks. $2\sigma_y^N$ is roughly the maximum
peak amplitude change that noise can cause. So, if a
peak in a total map whose distance
to the nearest peak in the corresponding clean SZ map is less than one
FWHM of the filter and its value is in the $2\sigma_y^N$ range of the
real value, we classify it as real. (3) We 
estimate the accuracy of $y$ peak CDF by comparing clean SZ maps,
noise maps and total maps (fig. \ref{fig:finalpeak}). For signal peaks 
with $y\gg \sigma_y^N$, the signal peaks remain to be peaks in the total
map. Noise mainly changes the
amplitude of  the $i$th real signal  from $y_i$ to
$y_p=y_i+y^N$ in the total map. Here, $y^N$ is the value of the noise
and $y_p$ is the value in the total map. Since noise is Gaussian, we
know the distribution of $y^N$, which can be described by
$P(y,\sigma_y^N)$, the probability  for the Gaussian noise with
dispersion $\sigma_y^N$ to have value bigger than $y$.   
Then we can related the CDF of  peaks in clean maps to the one of
total maps by
\begin{equation}
\label{eqn:ntot}
N_{tot}(y>y_p)=\sum^N_i P(y_p-y_i,\sigma_y^N).
\end{equation}
This relation gives a good fit in the $y\gg\sigma_y^N$ regime. 
For signals with $y\sim \sigma_y^N$,  noise changes 
both $y$ value and positions of some peaks, makes some peaks disappear
and  introduces a large fraction of false peaks. The sum of
$N_{tot}(y>y_p)$ in eqn. (\ref{eqn:ntot}) and $N_{noise}(y>y_p)$ considers
the effect that noise introduces false peaks and we expect that it would
give a good fit  in the region where $y\sim 2 \sigma_y^N$. We show the
modelled CDF of peaks in total maps following above procedures in
fig. \ref{fig:recover}. The result is well fitted in the range $y>2
\sigma_y^N$ with better than $30\%$ accuracy.

The optimal survey should find clusters as quickly as possible and
measure the $y$ parameter and peak CDF as accurately as possible.
We find  that the optimal
scan rate for cluster searching is about $280\ 
{\rm hours}\ {\rm deg}^{-2}$. At this rate, the rate of
cluster detection is  about $1$ every $7$ hours allowing for a false positive
rate $20\%$. The measured 
cluster CDF $N(y>y_p)$ is accurate to $30\%$ up to $N\simeq
60\ {\rm deg}^{-2}$. We recall 
that the optimal scan rate for SZ power spectrum measurement is
about $10\ {\rm hours}\ {\rm deg}^{-2}$. At this rate, the cluster
detection rate is about $1$ cluster every $60$ hours. We can  consider
the effect of cosmology on our estimation. A smaller $\sigma_8$
reduces signals and therefore the cluster detection rate. We assume
the extremest dependence of the $y$ parameter on $\sigma_8$, namely,
$y\propto \sigma_8^3$ as predicted by \citet{Zhang01a}. For
$\sigma_8=0.9$, the detection rate decreases by a factor of four. But
the optimal scan rate remains approximately the same.

   At this optimal scan rate, in a $1000$ hour AMIBA
survey, several hundred  clusters can be
found. Comparing the SZ cluster counts to Press-Schechter predictions
or X-ray surveys (\S \ref{sec:application}) allows us to reconstruct
the thermal history of the 
IGM and non-gravitational heating, which is expected to arise from
galaxy formation feedback \citep{Pen99}.

\section{Conclusion}
\label{sec:conclusion}

We have performed the largest high resolution SZ simulations to date, and
analysed the results.  We found further increases in the power
spectrum relative to previous simulations, and significantly more
small scale structures.  This trend is confirmed in Press-Schechter
estimates, and suggests nominally increasing power spectra.  The small
structure  behavior may not be a robust prediction,
since non-gravitational effects will significantly modify those scales.
We have examined the skewness and kurtosis on the sky maps, found
a strong non-Gaussianity on subdegree scales and confirmed the
log-normal distribution found in previous studies.  The Gaussian
estimates of power spectrum sample variances are less severely
affected, due to the averaging of many patches by each Fourier modes. 
But its effect on the power spectrum error analysis is significant.

We simulated SZ observations with AMIBA, and
analyzed the sensitivity for different scan rates.  In a $1000$ hour
survey, the optimal strategy for power spectrum estimations is to
scan several hundred square degrees.  The SZ angular power spectrum
measured in such survey can be determined to an accuracy of
$\sim 40\%$ over a range of $2000 \lesssim l \lesssim 5000$.  A cross
correlation with SDSS should allow an accuracy of $20\%$ in the cross
correlation measurement, which suggests that the time resolved
measurement of the pressure power spectrum is then possible.  This
scan rate results in a low  detection rate of 
clusters of galaxies, approximately one per $60$ hours with a false
positive rate of 20\%.  For the purpose of cluster search, the
optimal scan rate 
is around $280$ hours per square degree, which could find $1$ cluster
every $7$ hours, with the accuracy of $30\%$ in the cluster $y$ measurement
and $30\%$ in the cluster CDF up to $N(y>y_p)\sim 60\  {\rm deg}^{-2}$.

The predicted SZ power spectrum is consistent with recent indications
from the CBI experiment \citep{Mason01a,Sievers01} and the BIMA
upper limit ($95\%$ confidence). But it is higher than the BIMA
$1$-$\sigma$ result.  This may be a first 
indication of IGM non-gravitational feedback. Future blank sky surveys
with data analysis  considering actual SZ non-Gaussianity 
will provide us with a quantitative understanding of the thermal
history of the universe.

\acknowledgments
{{\it Acknowledgments}:
We thank Uros Seljak for the code to generate 2D
SZ maps and lots of helpful suggestions. We thank Josh Frieman,
Scott Dodelson and Albert Stebbins for the discussion of SDSS.}

\end{document}